\documentclass{ws-ijmpe}
\usepackage[super,compress]{cite}
\begin{document}

\markboth{Authors' Names}{Instructions for typing manuscripts (paper's title)}

\catchline{}{}{}{}{}

\title{Charged particle production in $pp$ collisions at $\sqrt{s}$=8, 7 and 2.76~TeV at the LHC$-$a case study}

\author{R.~Aggarwal}           
\address{Department of Technology, Savitribai Phule Pune University, Pune-411 007, India.\\
ritu.aggarwal1@gmail.com}

\author{M.~Kaur}
\address{Physics Department, Panjab University, Chandigarh, India\\
manjit@pu.ac.in \footnote{corresponding author}}

\maketitle

\begin{history}
\received{Day Month Year}
\revised{Day Month Year}
\end{history}

\begin{abstract}
We analyse the charged${\text -}$particle multiplicity distributions measured by the ALICE experiment, over a wide pseudorapidity range, for $pp$ collisions at $\sqrt{s}$=8,\,7\,and\, 2.76~TeV at the LHC.~The analysis offers an understanding of particle production in high energy collisions in the purview of a new distribution, the shifted Gompertz distribution.~Data are compared with the distribution and moments of the distributions are calculated.~A modified version of the distribution is also proposed and used to improve the description of the data consisting of two different event classes; the inelastic and the non${\text -}$single${\text -}$diffractive and their subsets in different windows of pseudorapidity, $\eta$.~The distribution used to analyse the data has a wide range of applicability to processes in different fields and complements the analysis done by the ALICE collaboration in terms of various LHC event generators and IP-Glasma calculations.  

\end{abstract}

\ccode{PACS numbers: 5.90.+m, 12.40.Fe, 13.66.Bc}


\section{Introduction}
With the advent of the highest energy particle accelerators such as the Large Hadron Collider (LHC) or Relativistic Heavy Ion Collider~(RHIC), several new and challenging searches have become possible.~Complexity of particle collisions at such energies and the occurence of possible new phenomena is expected to modify the global characteristics of collisions at these colliders.~Charged-particle multiplicity is one of the important global characteristics in high energy particle interactions which can serve to reveal the interaction${\text -}$dynamics.~This is one variable which has been studied for all kinds of particle and heavy${\text -}$ion interactions and measured in all types of particle detectors, for every collision energy.~Several important conclusions have followed from these studies, related to the average number of particles produced, their charge and momentum measurements.~At the high energy colliders, the interactions have a dominant contribution from the soft processes, involving low momentum${\text -}$transfer and are described by non${\text -}$perturbative Quantum Chromodynamics~(QCD).~The particle production can then be studied using phenomenological and statistical modelling.~However, as the interaction energy increases, the particle production also receives contribution from hard scattering, to be treated within the perturbative QCD regime.~Studies of charged-particle production are refining the understanding of global properties of proton ${\text -}$proton collisions at the LHC. 

The number of particles produced in a high energy collision and the governing production mechanism have been studied in terms of several theoretical, phenomenological, and statistical models.~A long list of such models exist in literature, a few more recent ones can be found in Refs.~[\refcite{a1},~\refcite{a2},~\refcite{a3},~\refcite{a4},~\refcite{a5}].~The negative binomial distribution~(NBD) [\refcite{a6}] is one such distribution which has been tested with every collision data, very successfully.~Each of these models derive from different assumptions and has shown premise of successful explanation of the data from different experiments.~Further extrapolations of these models have been done to make predictions for future experiments.~Two years back we proposed to anlyse the high energy collision data in terms of shifted Gompertz distribution.~The details of the distribution and justification for its applicability are published in the Ref.~[\refcite{a7}].~For the sake of completeness, in the following section a very brief introduction of this distribution is given.~Our first attempt to investigate the multiplicity distributions of charged particles produced in $e^{+}e^{-}$, $\overline{p}p$ and $pp$ collisions at different center of mass energies, with data in different phase space regions, produced very good results.~This motivated us to subsequently extend its validity to other collision data.~We have also performed comprehensive analysis of hadron${\text -}$nucleus data from fixed target experiments at relatively low energies~[\refcite{a8}]. 

As another case study, in this paper the shifted Gompertz distribution with non${\text -}$negative fit parameters, identified with the scale and shape parameters, is used for studying the charged multiplicity distributions of $pp$ collisions in the ALICE detector  at the LHC at three different centre${\text -}$of${\text -}$mass energies Refs.~[\refcite{ALICE1}, \refcite{ALICE2}].~It is worthwhile to mention that the ALICE collaboration has analysed this data by using NBD.~The important reason to consider this data is the fact that the data was recorded by the collaboration in very wide range of pseudorapidity and in different phase space windows; $|\eta|$=0.5, 1.0, 1.5, 2.0, 2.5, 3.0, 3.4, 5.0.~In addition the data are also classified into two different classes of events; the inelastic~(INEL) and the non${\text -}$single${\text -}$diffractive~(NSD).~The inelasic sample is further categorised as inelastic~(INEL) and inelastic$>$0~(INEL$>$0).~The difference being the inclusion of data for charged multiplicity $n${\text =}$0$ point.~This way 22 data sets are available.~It is well established that for the full phase space, energy${\text -}$momentum conservation rule strongly influences the multiplicity distribution.~The distribution in a restricted rapidity window, however, is less prone to such constraints and thus can be expected to be a more sensitive probe of the underlying dynamics of QCD, as inferred in Refs.~[\refcite{a11},~\refcite{a12}].~The high energy interactions, dominated by soft processes which involve small${\text -}$momentum${\text -}$transfer, are useful to study QCD in the non${\text -}$perturbative regime, and to constrain phenomenological models.~They are also important for the understanding of backgrounds for measurement of hard and rare processes.

\section{Parameterizing multiplicities with the shifted Gompertz Probability distribution Function (PDF)}
The shifted Gompertz distribution [\refcite{a13}] is a distribution of the largest of two independent random variables one of which has an exponential distribution with parameter $b$ and the other has a Gumbel distribution, also known as log${\text -}$Weibull distribution, with parameters $t$ and $b$.~Several of its properties have been studied with application to problems in computer science, statistics and applied mathematics [\refcite{a14},~\refcite{a15},~\refcite{a16}].~As the particle production in high energy collisions follows certain probability laws, we thought that it might be interesting to study the statistical phenomena in high energy physics in terms of this distribution.~The shifted Gompertz distribution with non${\text -}$negative fit parameters identified with the scale and shape parameters, can in this way be used for studying the distributions of particles produced in collisions at colliders.

For a sample of high energy collisions producing $n$ particles on the average, the probability density function~(PDF) is defined by two non${\text -}$negative parameters, $b$, the scale parameter and $t$ the shape parameter.~Following equations define the PDF and the mean of the distribution;
\begin{equation}
P\big(n|b,t\big) = b e^{-bn}e^{-\big(t e^{-bn}\big)}\big[1+t(1-e^{-bn}\big)\big]\hspace{0.3cm} for\hspace{0.15cm} n > 0 \,. \label{SGD}
\end{equation}
Mean of the distribution:
\begin{equation}
 \Big(-\frac{1}{b}\Big)\Big(E\big[ln(\zeta)\big]-ln(t)\Big) \hspace*{0.3 cm} where \hspace*{0.2 cm} \zeta = t e^{-bn}
􏰌\,. \label{SGDm}
\end{equation}
and
\begin{equation}
 E\big[ln(\zeta)\big] = \Big[1 + \frac{1}{t}\Big]\int_{0}^{\infty}e^{-\zeta}\big[ln(\zeta)\big]d\zeta 
 - \frac{1}{t}\int_{0}^{\infty}\zeta e^{-\zeta}\big[ln(\zeta)\big]d\zeta\
􏰌\,. \label{SGDm2}
\end{equation}
\normalsize
where $b \geq 0$ and $t \geq 0$ are respectively the scale and shape parameter. 
Although shifted Gompertz~(SG) distribution has been studied recently, with very good results in hadronic and leptonic interactions, in this paper the aim is to extend its validity to higher energy and to different types of events, as explained above.~The mean of the distribution given in~Eq.~(\ref{SGDm}) is expected to approximate the mean of the experimental distribution,$\langle{n}\rangle=\sum\limits_{n=0}^{n_{max}} {nP(n)}$. 

\subsection{Parameterization with superposition of two shifted Gompertz PDFs}
It is well established that at high energies, charged particle multiplicity distribution in full phase space becomes broader than a Poisson distribution.~This behaviour was successfully explained by describing the distribution in terms of negative binomial function (NB), whereby the distribution is defined in terms of two free parameters [\refcite{a6}].~One parameter gives the average charged multiplicity $\langle{n}\rangle$ and the other parameter $k$ relates to the dispersion of the distribution as given below:
\small
\begin{equation}
P(n|\langle{n}\rangle,k)=\frac{\Gamma(n+k)}{\Gamma(n+1)\Gamma(k)}\frac{(\langle{n}\rangle/k)^n}{(1+\langle{n}\rangle/k)^{n+k}}
􏰌\,. \label{NBD} 
\end{equation}
\normalsize
where $k$ is related to the dispersion $D$ by
\small
\begin{equation}
\frac{D^2}{\langle{n}\rangle^2}=\frac{1}{\langle{n}\rangle}+\frac{1}{k}􏰌\,. \label{DISP}
\end{equation} 
\normalsize
The success of NBD was phenomenal, in the sense that it could describe almost all the data available at different energies.~Until, the NB violation became imminent from the measurements done by the UA1 collaboration [\refcite{a17}] and the UA5 collaboration~[\refcite{UA5},~\refcite{a18}] on the $\overline{p}p$ collisions at 540~GeV, showed a shoulder structure in the multiplicity distribution.~A possible change in particle production dynamics at high energies was indicated.~To explain this NB violation, C. Fuglesang [\refcite{a19}] suggested that the violations may be on account of the presence of more than one class of events in the collision data.~He then proposed to account for the change in shape of the distribution by the weighted superposition of two classes of events; soft events (events without mini${\text -}$jets) and semihard events (events with mini${\text -}$jets), the weight $\alpha$ being the fraction of soft events.~The multiplicity distribution of each class following the NBD.~This idea was subsequently implemented by Giovannini et al in  analyses at high energies [\refcite{Gov}] and successfully described the multiplicity distributions for collision data from various colliders including Tevatron, LEP and LHC.~In this approach, the multiplicity distribution depends on five parameters; two for each of the superimposed distributions and the fraction $\alpha$ of soft events.

The appearance of substructures in multiplicity distributions becomes more predominant in the wider phase space regions.~In the present paper, when the data were analysed with the shifted Gompertz distribution, we encounter the trends similar to the NB.~Hence we adopted the two${\text -}$component approach and fitted the two weighted SG distributions as below;
\begin{equation}
P_{n(2SGD)} = \lambda [\alpha P_{n(SGD)} + (1 - \alpha) P_{n(SGD)}]􏰌\,. \label{2SGD}
\end{equation}
With this, the fits to the data improved manifold.~For INEL and NSD class of events the value of $P(n=0)$ is non-zero which occurs due to the limited  $\eta$ acceptance of the detector.~And this type of function~(NB or SG) is not meant to describe the value $P(n=0)$.~Therefore this point was excluded from the fit and subsequently from the calculation of moments.~An overall normalization factor $\lambda$ was introduced, as a free parameter, is applied to account for this.

\subsection{Moments of multiplicity distributions}
For a multiplicity distribution, the normalised moments $C_{q}$, normalised factorial moments $F_{q}$, normalised factorial cumulants $K_{q}$ and ratio of the two $H_{q}$ moments are defined as;

\begin{equation}
C_{q} = \frac{\sum_{n=1}^{\infty}n^{q}P(n)}{(\sum_{n=1}^{\infty}nP(n))^q}􏰌\,. \label{Cq}
\end{equation}
\begin{equation}
F_{q} = \frac{\sum_{n=q}^{\infty}n(n-1).......(n-q+1)P(n)}{(\sum_{n=1}^{\infty}nP(n))^q}􏰌\,. \label{Fq}
\end{equation}
\begin{equation}
K_{q} = F_{q}-\sum_{m=1}^{q-1}\frac{(q-1)!}{m!(q-m-1)!}K_{q-m}F_{m}􏰌\,. \label{Kq}
\end{equation}
\begin{equation}
H_{q} = K_{q}/F_{q}􏰌\,. \label{Hq}
\end{equation}

Where $P(n)$ is the $n{\text -}$charged particle probability and $q$ is the rank of the moment.
In the specific case of the shifted Gompertz distribution, the normalized moments ($C_q$) and normalized factorial moments ($F_q$) are defined as following with $n$ as values of $X$:
\begin{equation}
C_q = \frac{E[X^{q}]}{(E[X])^{q}} \>\>\>\>and\>\>\>\> F_q = \frac{E[(X)(X-1)(X-2)....(X-(q-1))]}{(E[X])^{q}}􏰌\,. \label{Cq1}
\end{equation} 

$q$ is a natural number ranging from 1 to $\infty$.~The Mean value (E[X]) of Shifted Gompertz distribution is given by 
\begin{equation}
E[X] = \frac{1}{b}( \gamma + \ln t + \frac{1 - e^{-t}}{t} + \Gamma[0,t])􏰌\,. \label{EX1}
\end{equation}
\begin{equation}
E[X^{2}] = \frac{2}{b^{2} t} \Big(\gamma + \Gamma[0, t] + t^{2}\>_3F_3[\{1,1,1\},\{2,2,2\},-t] + \ln t\Big)􏰌\,. \label{EX2} 
\end{equation}
The higher order moments~($C_{q}$) can be found by using the moment generating function of the Shifted Gompertz distribution[[$b$,$t$],$m$] 
\begin{equation}
e^{-t} - (1 + \frac{m}{bt}) t^{\frac{m}{b}}(\Gamma[1-\frac{m}{b}] - \Gamma[1-\frac{m}{b},t])􏰌\,. \label{et}
\end{equation}
\begin{equation}
\begin{aligned}
E[(X)(X-1)] ={}&\frac{2}{b^{2}t}\Big(\gamma + \Gamma[0,t] + t^{2} \>_3F_3[\{1,1,1\},\{2,2,2\},-t] + \ln t\Big) - \\
&\frac{\Big(1-e^{-t} + t (\gamma + \Gamma[0,t] + \ln t)\Big)}{bt}
\end{aligned}􏰌\,. \label{EXX}
\end{equation}

The higher order factorial moments~($F_q$) can be found by using the generating function of Shifted Gompertz Distribution[[$b$,$t$],$m$] 
\begin{equation}
e^{-t} - t^{\frac{\ln m}{b}} (\Gamma[1-\frac{\ln m}{b}] - \Gamma[1-\frac{\ln m}{b},t]) (1  + \frac{\ln t}{b t})􏰌\,. \label{ett}
\end{equation}
where\\
(i) $\gamma$ $\approx$ 0.5772156 is the Euler constant (also referred to as Euler-Mascheroni constant).
\\
(ii) $\Gamma$[s] the Euler Gamma function and $\Gamma[s,n]$ the incomplete Gamma function are defined as;
\begin{equation}
\Gamma[s] = \int_{0}^{\infty} m^{s-1} e^{-m}dm \hspace{ 1cm} \Gamma[s,n] = \int_{n}^{\infty}  m^{s-1} e^{-m}dm
\end{equation}
(iv) $_3F_3$ is a Generalized Hypergeometric function. 
\begin{equation}
_3F_3[\{1,1,1\},\{2,2,2\},-t] = \sum_{k=1}^{\infty} \frac{(-1)^{k+1} t^{k+1}}{k! k^{2}} 
\end{equation}
\subsection{Charged particle multiplicities analysed}
Charged particle multiplicity distributions in $pp$ collision data at $\sqrt{s}$=\,8,\,7 and 2.76~TeV, collected by the ALICE collaboration [\refcite{ALICE1},~\refcite{ALICE2}] are  analysed.~Events are classified into three classes depending upon the experimental requirements.~These are;\\
i) class${\text -}$I:  all inelastic events~(INEL).\\ 
ii) class${\text -}$II: requires the presence of at least one charged particle (tracklet) in the region $|\eta|<$1.0 in addition to the INEL condition. \\
iii) class${\text -}$III:  Non${\text -}$single${\text -}$diffractive~(NSD) events.\\
~For each class, multiplicity distributions at all energies, are studied in two groups:\\
i) central region with pseudorapidity $|\eta|<$ 0.5, 1.0 and 1.5.\\
ii) forward region with pseudorapidity $|\eta|<$ 2.0, 2.4, 3.0, 3.4 and for $\eta$= -3.4$-$5.0.  
\section{Results and discussion}
The experimental charged multiplicity distributions are studied with the PDF of shifted Gompertz distribution (SGD) and a convolution of two SGDs~(henceforth referred to as 2SGD).~The convolution is done by superposing two weighted SGDs.~The SGD is a two${\text -}$parameter distribution, namely scale and shape parameters.~As a result the weighted superposition has five free parameters including the weight factor.~The PDFs are calculated by using Eq.(\ref{SGD}) and Eq.(\ref{2SGD}), normalised and matched with the data by carrying out minimum $\chi^{2}$ fits using ROOT6.18.

In the present analysis, due to large number of data subsets, total number of figures produced is more than 100.~Thus we shall be including only few figures of each type of results to avoid the multitude of similar figures leading to cluttering.  
\subsection{Multiplicity distributions at different energies}
\begin{figure}[th]
\centerline{\includegraphics[scale=0.38]{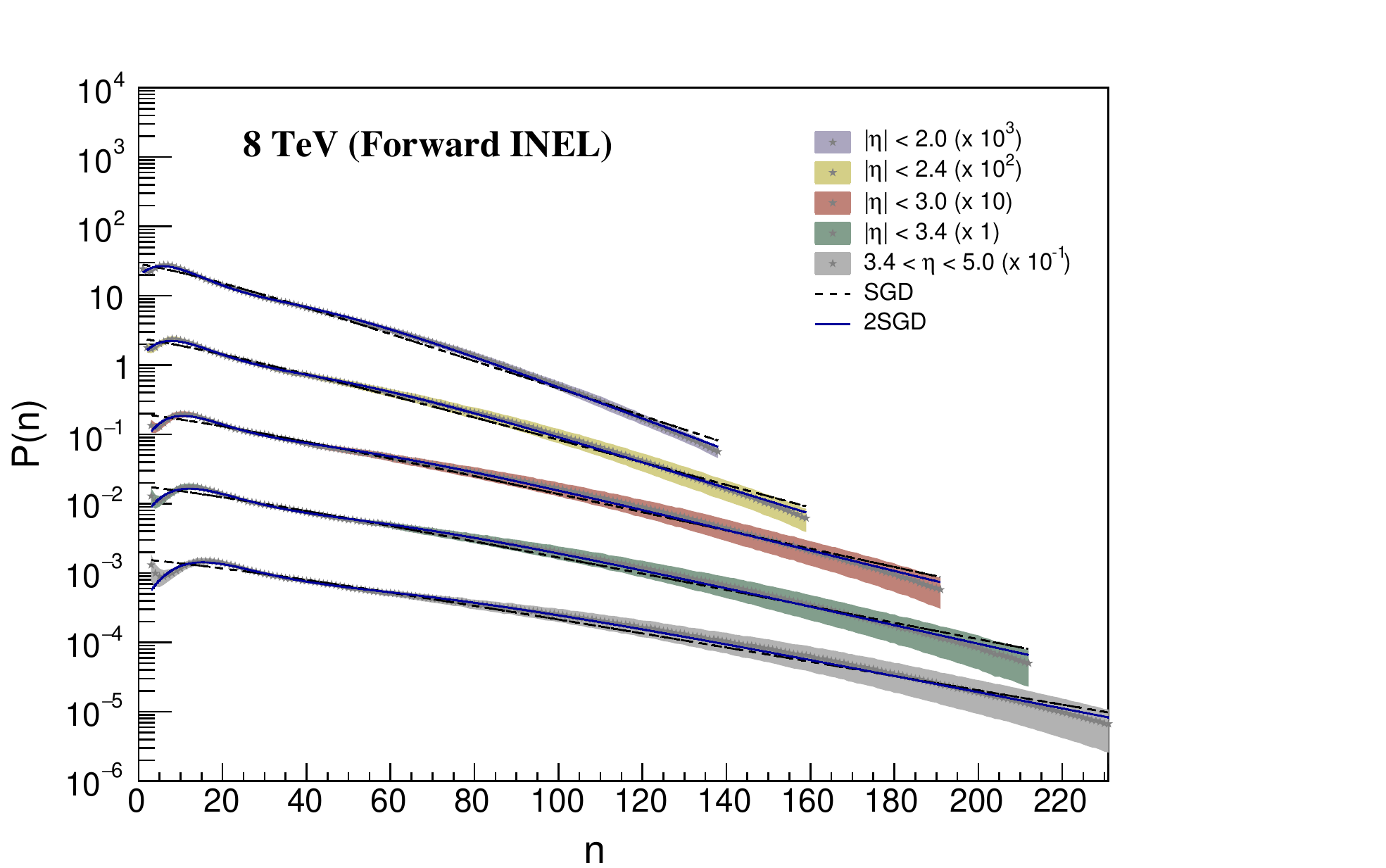}\includegraphics[scale=0.38]{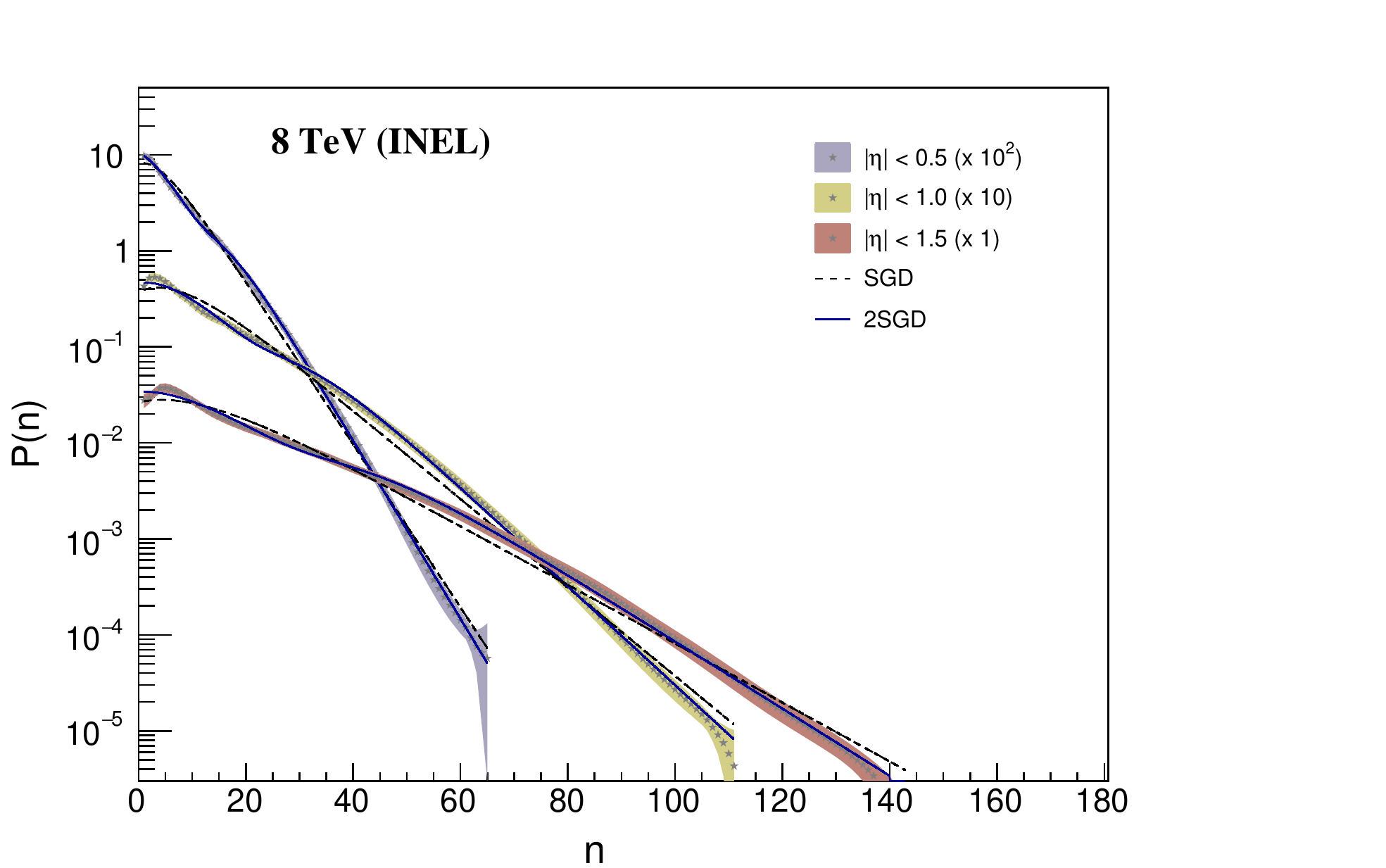}}
\centerline{\includegraphics[scale=0.38]{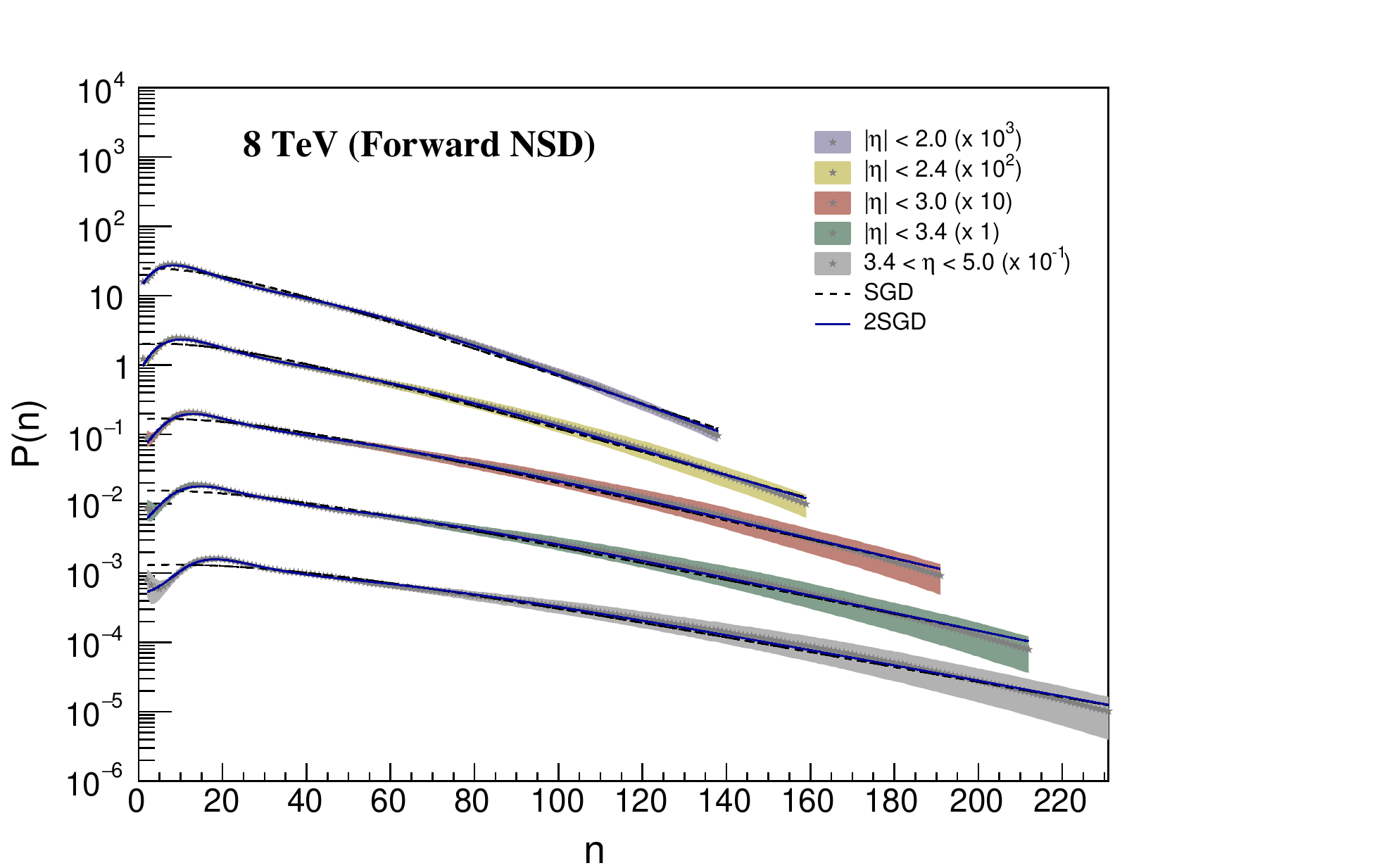}\hspace{0.01cm}\includegraphics[scale=0.38]{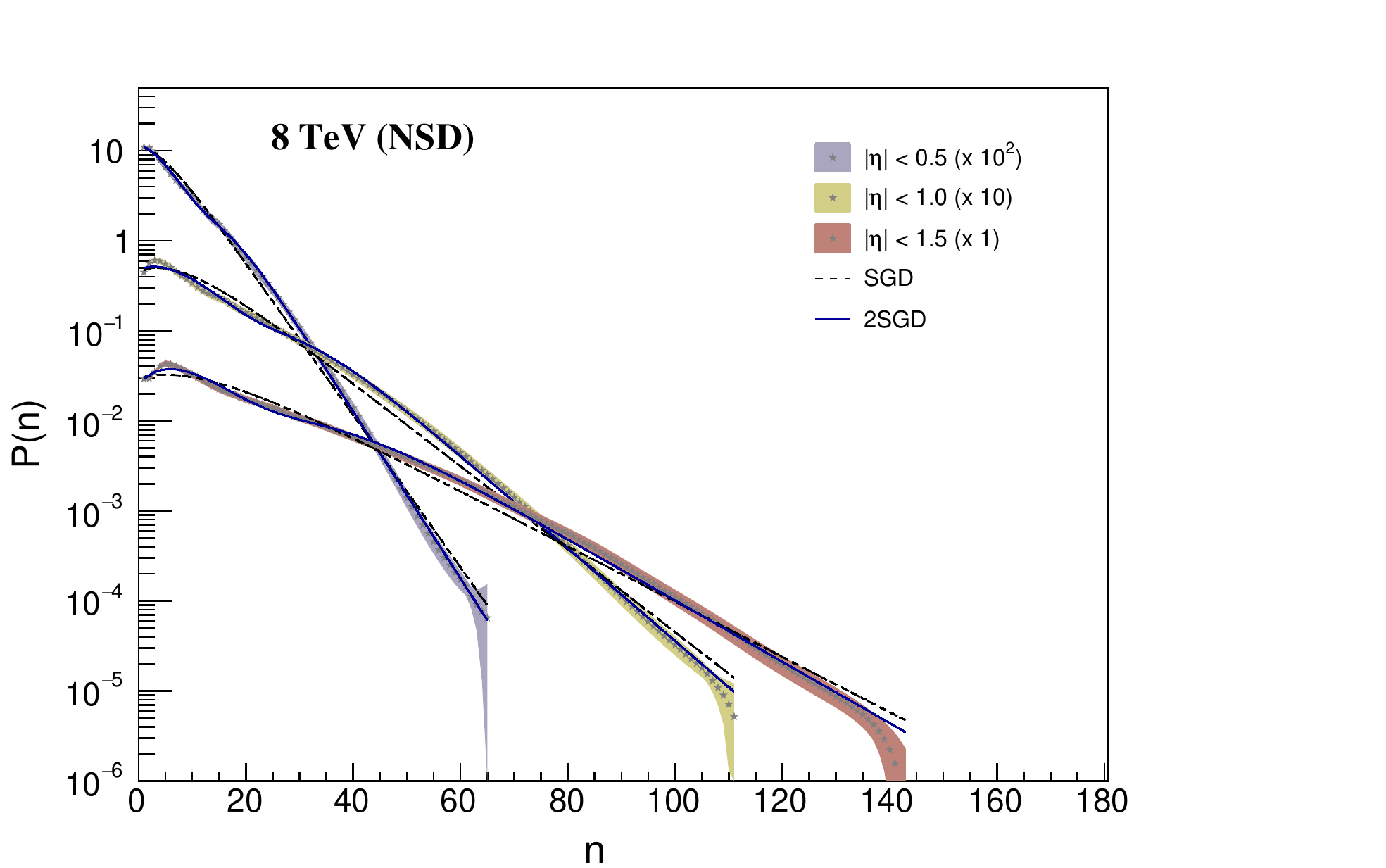}}
\caption{Probability distribution of $n{\text -}$charged particles in 8~TeV non-single-diffractive events~(left) in the forward region and in the central pseudorapidity region~(right) of the ALICE detector at the LHC.~Legends point to the data and the fit distributions, with shaded areas representing combined systematic and statistical uncertainties on the data.}
\label{fig8pn}
\end{figure}

\begin{figure}[th]
\centerline{\includegraphics[scale=0.38]{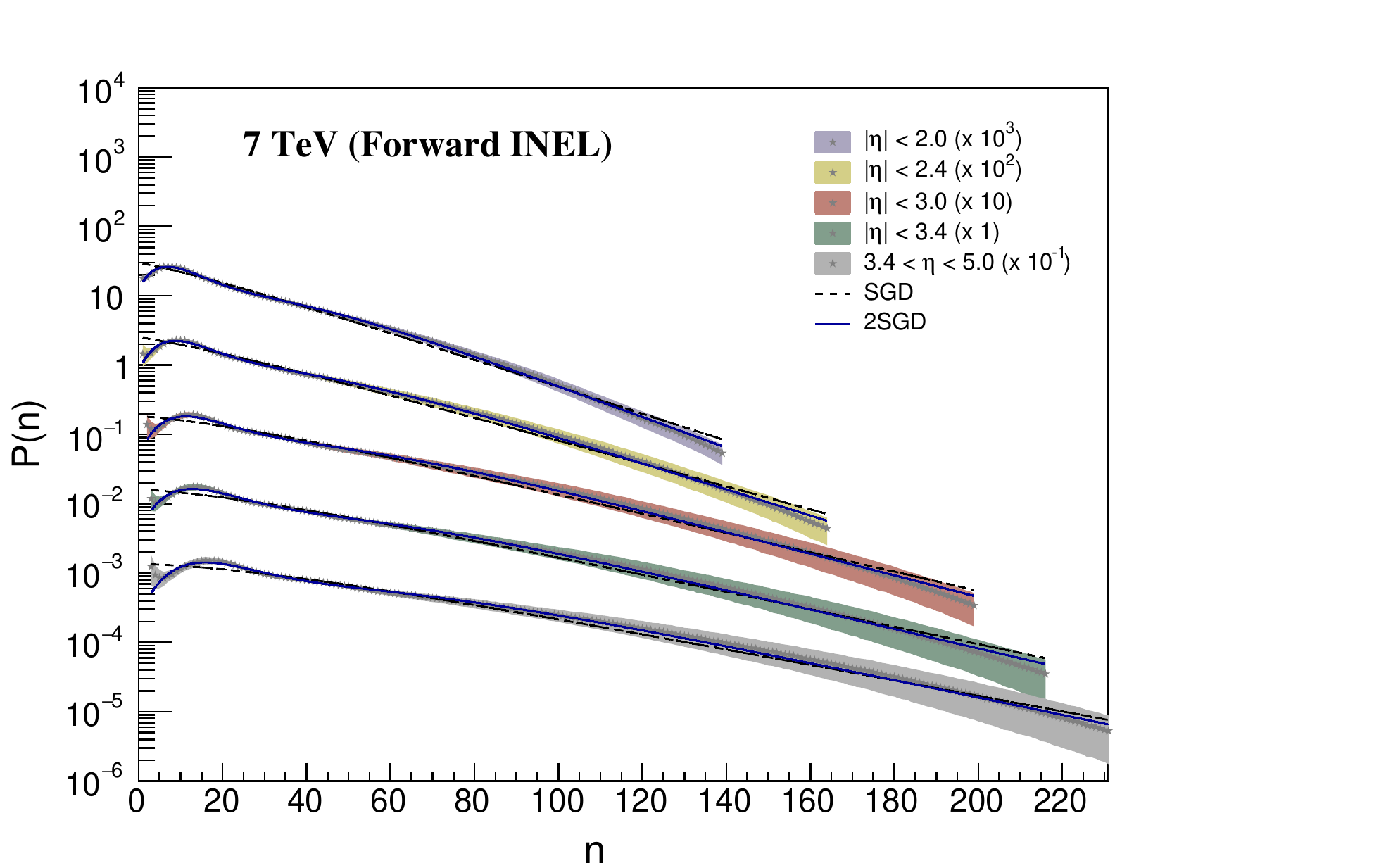}\includegraphics[scale=0.38]{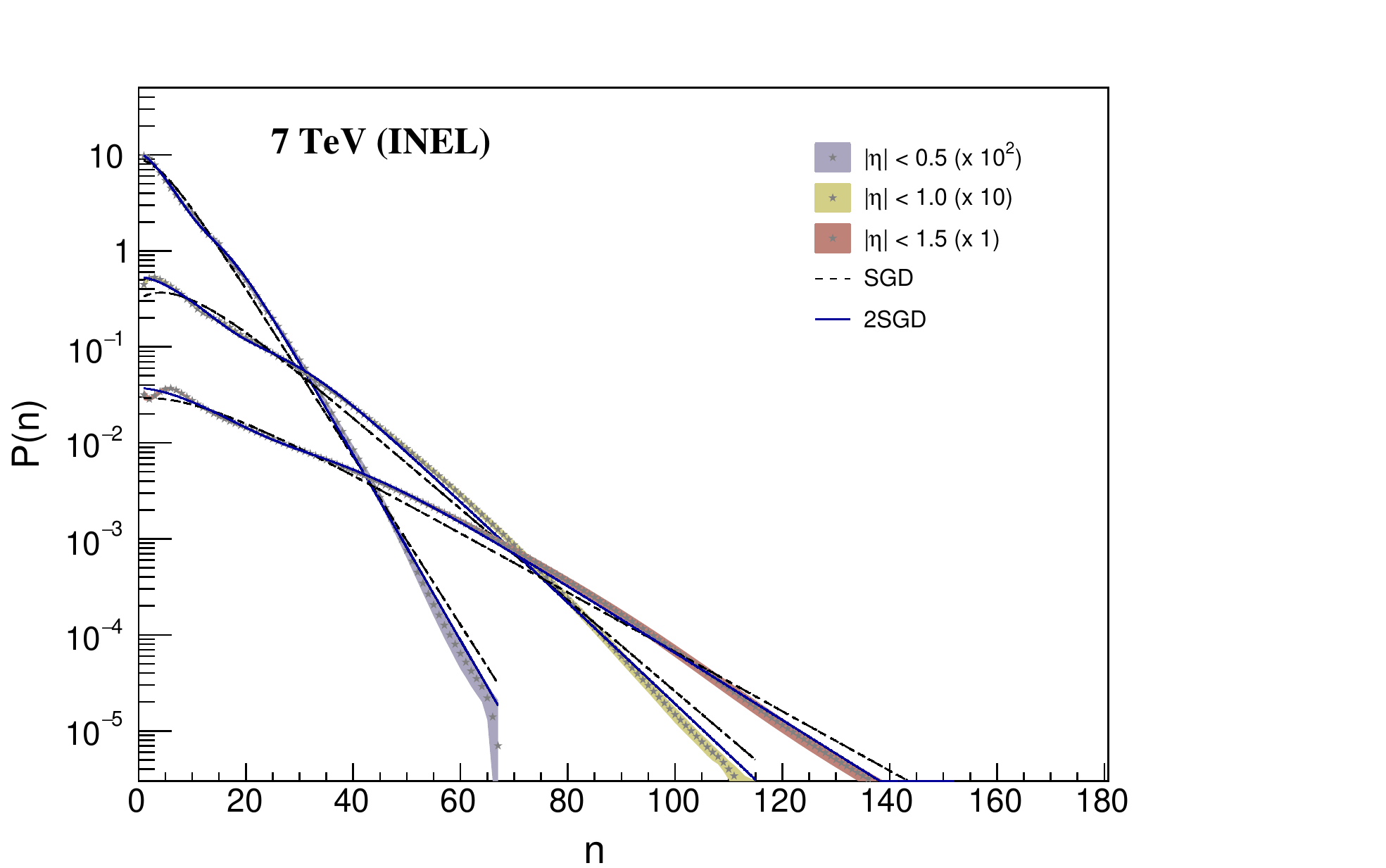}}
\centerline{\includegraphics[scale=0.38]{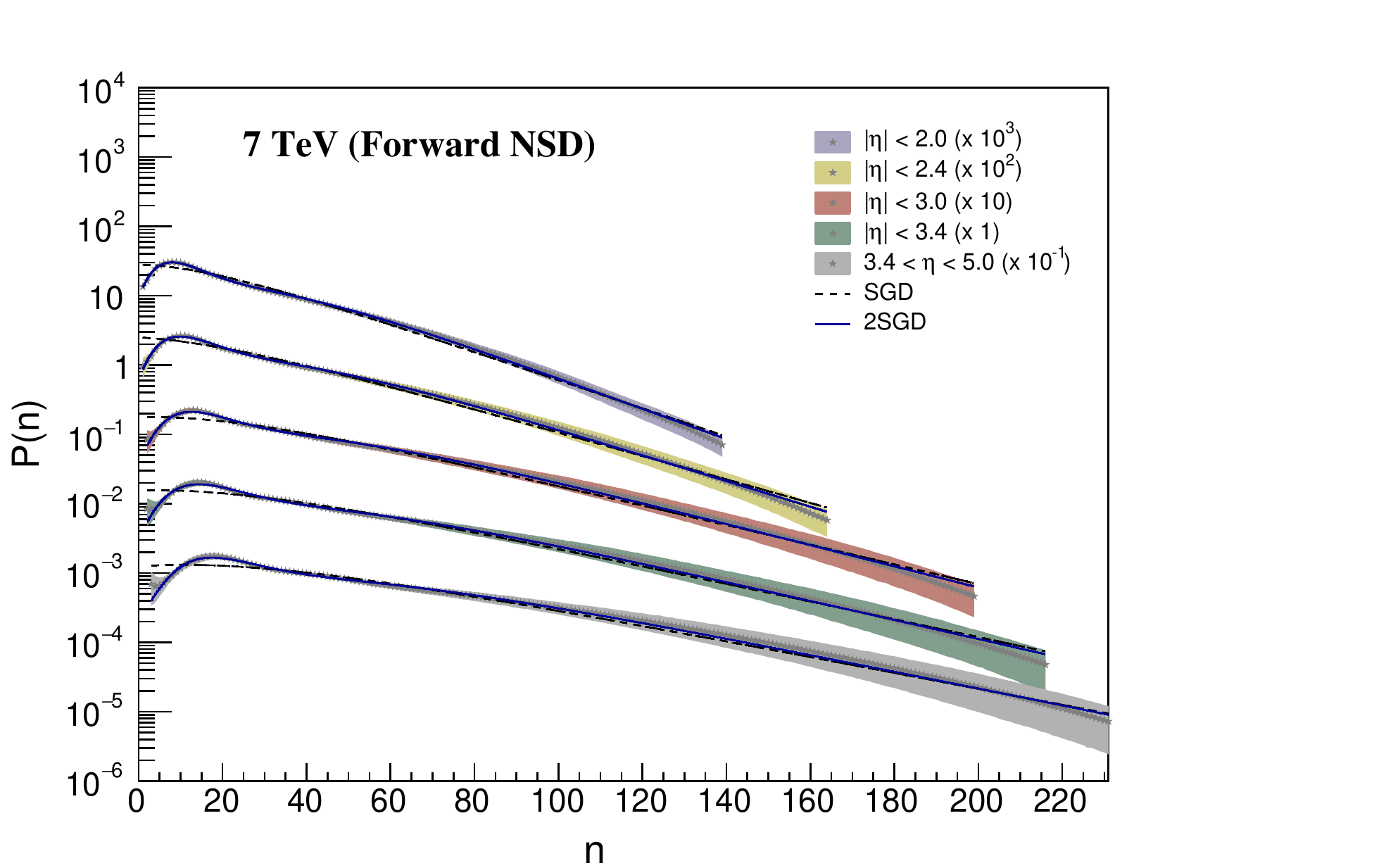}\includegraphics[scale=0.38]{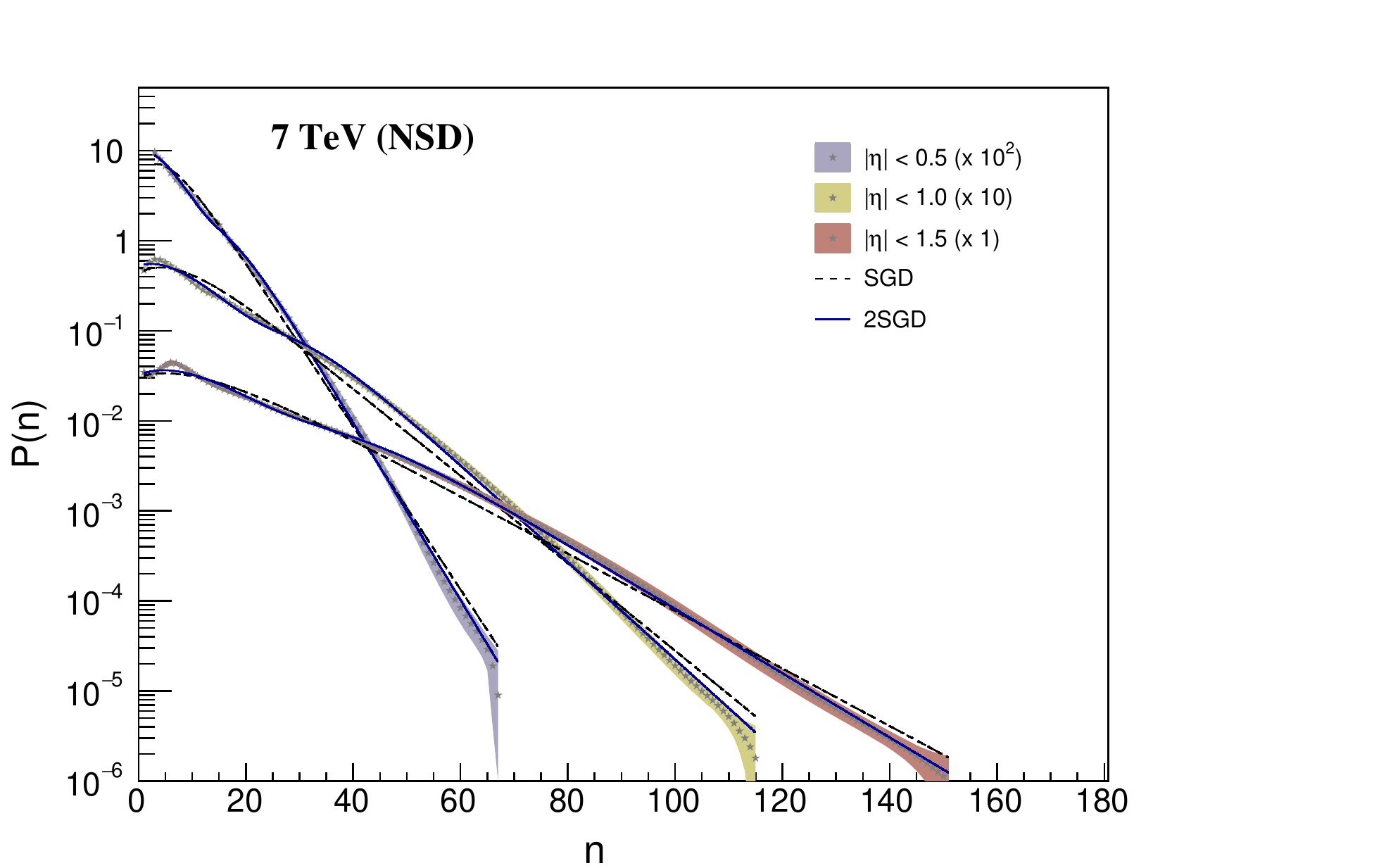}}
\caption{Probability distribution of $n{\text -}$charged particles in 7~TeV inelastic events~(left) in the forward region and in the central pseudorapidity region~(right) of the ALICE detector at the LHC.~Legends point to the data and the fit distributions, with shaded areas representing combined systematic and statistical uncertainties on the data.}
\label{fig7pn}
\end{figure}

\begin{figure}[th]
\centerline{\includegraphics[scale=0.38]{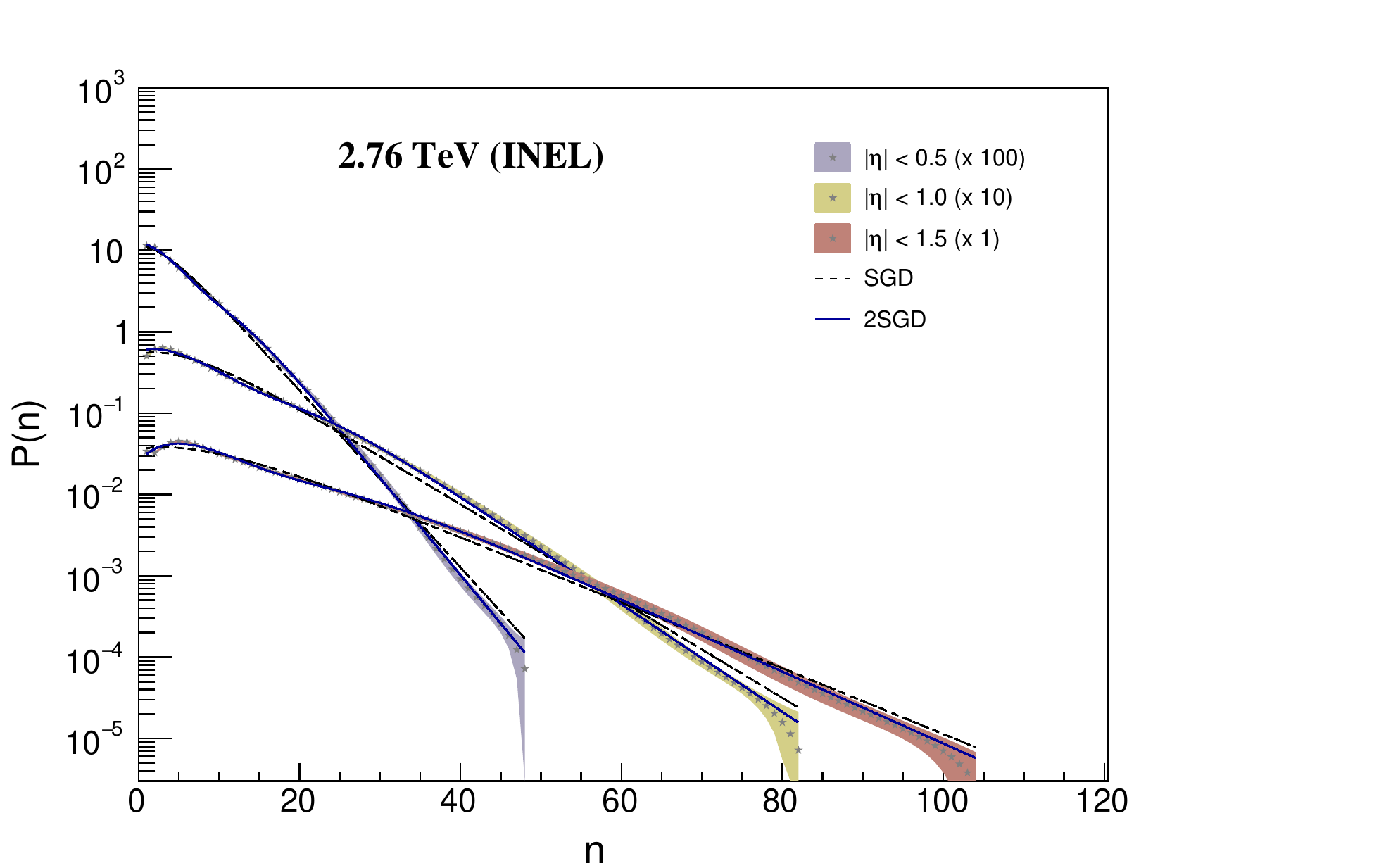}\includegraphics[scale=0.38]{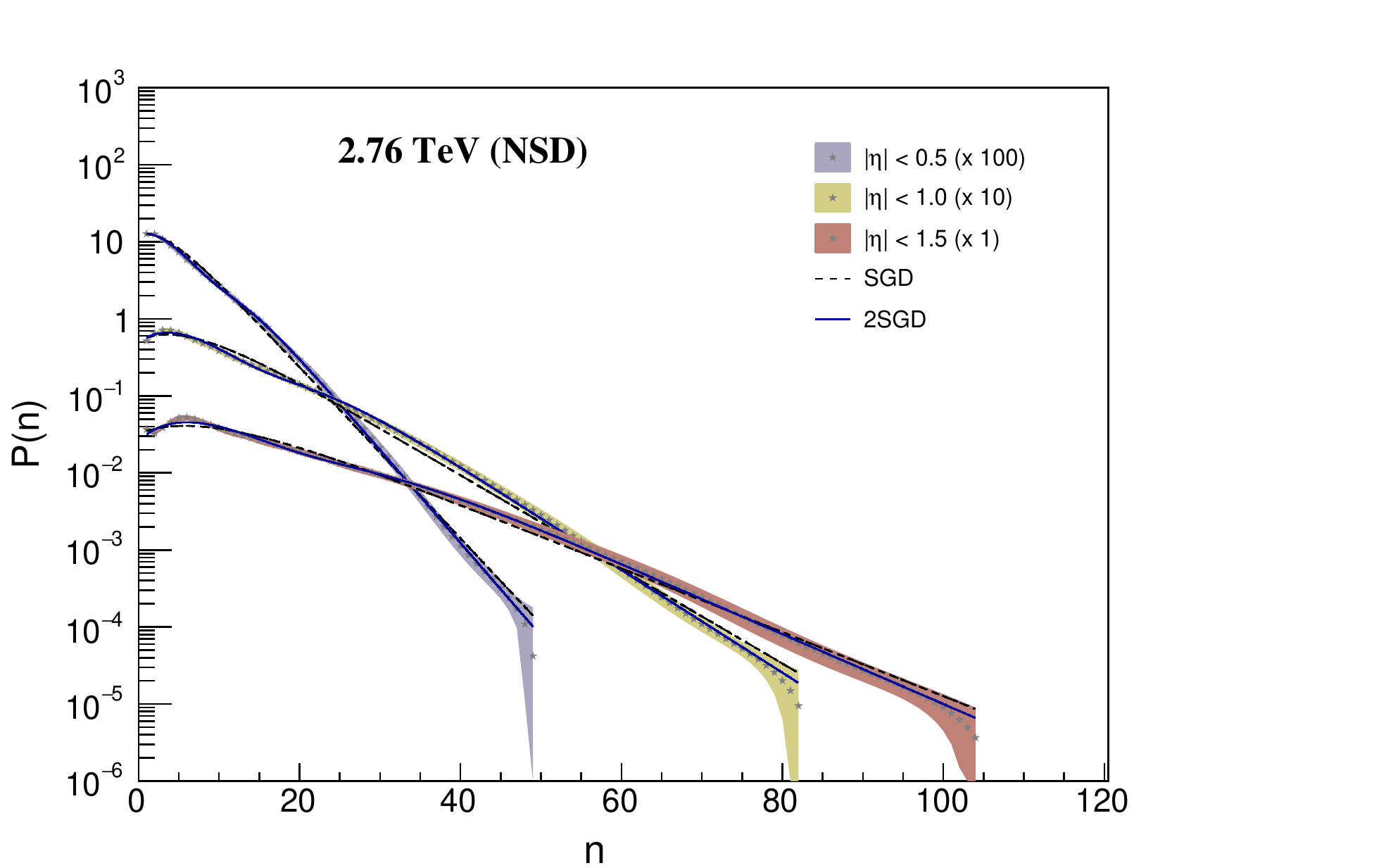}}
\caption{Probability distribution of $n{\text -}$charged particles in 2.76~TeV inelastic events~(left) in the forward region and in the central pseudorapidity region~(right) of the ALICE detector at the LHC.~Legends point to the data and the fit distributions, with shaded areas representing combined systematic and statistical uncertainties on the data.}
\label{fig276pn}
\end{figure}

The PDFs defined by Eq.(\ref{SGD}) and Eq.(\ref{2SGD}) are used to fit the experimental distributions of inelastic and non${\text -}$single${\text -}$diffractive events. Figure\ref{fig8pn} shows the shifted Gompertz function~(SGD) and the two${\text -}$component shifted Gompertz function~(2SGD) fits to the $pp$ collision data from the ALICE experiment at 8~TeV for the two classes of events and in different pseudorapidity windows.~Figures \ref{fig7pn}${\text -}$\ref{fig276pn} show similar plots for collision energies of 7~TeV and 2.76~TeV respectively.~For 2.76~TeV, data are available only in the intervals up to $|\eta|\leq$1.5.~It is observed that the shapes of the distributions reflect the fact that the high${\text -}$multiplicity tails of the distributions increase faster with increasing energy and with increasing pseudorapidity interval than the low multiplicity.~This feature also embeds the KNO scaling violation at these energies.

The difference between the multiplicity distributions for NSD and for INEL events becomes significant at low multiplicities, typically with $n<15$ as expected.

 \begin{table}[pt] 
  \scalebox{0.9}{ 
 \tbl{Ratio of $\chi^{2}/ndf$ for 2SGD fit and SGD fit for data at $\sqrt{s}$=8,\,7 and 2.76~TeV.} 
 {\begin{tabular}{@{}|c |c |c   |c   |c |@{}}  \toprule 
 \multicolumn{5}{|c|}{$(\chi^2/ndf)_{2SGD/SGD}$}\\\\
\hline 
&  $\eta$& 8 TeV  & 7 TeV & 2.76 TeV \\ \hline 
  \hline 
 & -0.5 $< \eta <$ 0.5 &       0.09 &       0.13  &           0.08  \\  
 INEL & -1.0 $< \eta <$ 1.0 &       0.14 &       0.13  &           0.09  \\  
 & -1.5 $< \eta <$ 1.5 &       0.16 &       0.13  &           0.17  \\  
\hline 
  INEL $>$ 0 & -1.0 $< \eta <$ 1.0 &       0.14 &       0.13  &           0.10  \\  
\hline 
  & -0.5 $< \eta <$ 0.5 &       0.09 &       0.15  &           0.09  \\  
 NSD & -1.0 $< \eta <$ 1.0 &       0.18 &       0.15  &           0.14  \\  
 & -1.5 $< \eta <$ 1.5 &       0.21 &       0.14  &           0.26  \\  
\hline  
  & -2.0 $< \eta <$ 2.0 &       0.11 &       0.06  & \\  
 & -2.4 $< \eta <$ 2.4 &       0.11 &       0.11  & \\  
 INEL & -3.0 $< \eta <$ 3.0 &       0.16 &       0.19   &\\  
 (Forward) & -3.4 $< \eta <$ 3.4 &       0.15 &       0.19   &\\  
 & -3.4 $< \eta <$ 5.0 &       0.16 &       0.21  & \\  
\hline 
  & -2.0 $< \eta <$ 2.0 &       0.09 &       0.04   &\\  
 & -2.4 $< \eta <$ 2.4 &       0.09 &       0.08   &\\  
 INEL $>$ 0 & -3.0 $< \eta <$ 3.0 &       0.12 &       0.11 &  \\  
 (Forward) & -3.4 $< \eta <$ 3.4 &       0.13 &       0.18 &  \\  
 & -3.4 $< \eta <$ 5.0 &       0.12 &       0.19   &\\  
\hline 
  & -2.0 $< \eta <$ 2.0 &       0.07 &       0.04 &  \\  
 & -2.4 $< \eta <$ 2.4 &       0.07 &       0.07 &  \\  
 NSD & -3.0 $< \eta <$ 3.0 &       0.09 &       0.12   &\\  
 (Forward) & -3.4 $< \eta <$ 3.4 &       0.10 &       0.13   &\\  
 & -3.4 $< \eta <$ 5.0 &       0.10 &       0.14 &  \\  
\hline 
 \end{tabular}} 
}
 \end{table}
Table~1 gives a comparative performance of fitting the two distributions, 2SGD versus SGD in terms of ratio of the $\chi^{2}/ndf$ at all the three energies, 8,\,7 and 2.76~TeV.~It may be observed that the ratio is $\ll$1 confirming that PDF of the convolution of two SGDs explains the data extremely well.~The fit parameters of 2SGD, the $\chi^{2}/ndf$ and the $p{\text -}$values for the data at all three energies are documented in tables~2${\text -}$4.~To avoid a very large number of tables, we are giving the tables of fit parameters only for the 2SGD at all energies.~Comparing the fittings, we find that the performance of SGD improves with the increase in collision energy.~On the other hand the 2SGD fits perfectly at almost all the energies, though the performance is much better in the forward region for both inelastic and NSD collisions.~The performance of 2SGD fit is poor for the inelastic coillsions in the range $|\eta|<$ 1 and $|\eta|<$ 1.5 at 7~TeV and in the range $|\eta|<$ 1.5 for 8~TeV, with large $\chi^{2}/ndf$ and $p{\text -}$values corresponding to CL\,$<$ 0.1$\%$. 

\begin{table}[pt]
 \begin{center}
 \scalebox{0.8}{
 \tbl{Fit parameter values, $\chi^{2}/ndf$ and $p{\text -}$values obtained from 2SGD for 8~TeV inelastic and single${\text -}$non${\text -}$diffractive events.}  
{\begin{tabular}{@{}|c c c c c c c c|@{}} \toprule
$\eta$ &   $b_{1}$ & $t_{1}$&   $\alpha$ & $b_{2}$&   $t_{2}$   & $\chi^2/ndf$ & $p$-$value$  \\ 
 \hline 
\hline 
 \multicolumn{8}{|c|}{INEL}\\ 
 \hline-0.5 $< \eta <$ 0.5  &     0.214 $\pm$    0.012&     0.381 $\pm$    0.108&     0.936 $\pm$    0.021&     0.214 $\pm$    0.008 &    31.098 $\pm$   11.987&       8.74 / 59 &     1.00  \\ \hline 
-1.0 $< \eta <$ 1.0  &     0.119 $\pm$    0.005&     0.635 $\pm$    0.075&     0.922 $\pm$    0.015&     0.118 $\pm$    0.003 &    38.261 $\pm$    9.022&      34.86 / 105 &     1.00  \\ \hline 
-1.5 $< \eta <$ 1.5  &     0.082 $\pm$    0.004&     0.568 $\pm$    0.091&     0.906 $\pm$    0.019&     0.080 $\pm$    0.002 &    29.250 $\pm$    6.788&      39.54 / 137 &     1.00  \\ \hline 
\hline 
 \multicolumn{8}{|c|}{INEL$>$0}\\ 
 \hline-1.0 $< \eta <$ 1.0  &     0.122 $\pm$    0.003&     0.401 $\pm$    0.034&     0.890 $\pm$    0.008&     0.112 $\pm$    0.001 &    19.680 $\pm$    1.189&     344.63 / 105 & $<$ 0.01 \\ \hline 
\hline 
 \multicolumn{8}{|c|}{NSD}\\ 
 \hline-0.5 $< \eta <$ 0.5  &     0.214 $\pm$    0.012&     0.454 $\pm$    0.074&     0.937 $\pm$    0.020&     0.213 $\pm$    0.008 &    31.800 $\pm$   11.436&      10.15 / 59 &     1.00  \\ \hline 
-1.0 $< \eta <$ 1.0  &     0.123 $\pm$    0.006&     0.784 $\pm$    0.073&     0.916 $\pm$    0.018&     0.117 $\pm$    0.002 &    36.582 $\pm$    8.132&      55.64 / 105 &     1.00  \\ \hline 
-1.5 $< \eta <$ 1.5  &     0.116 $\pm$    0.014&     1.223 $\pm$    0.151&     0.798 $\pm$    0.048&     0.078 $\pm$    0.001 &    15.986 $\pm$    4.224&      62.90 / 137 &     1.00  \\ \hline 
\hline \hline 
 \multicolumn{8}{|c|}{INEL (Forward)}\\ 
 \hline-2.0 $< \eta <$ 2.0  &     0.123 $\pm$    0.011&     1.139 $\pm$    0.126&     0.593 $\pm$    0.048&     0.052 $\pm$    0.001 &     4.460 $\pm$    0.906&      25.08 / 132 &     1.00  \\ \hline 
-2.4 $< \eta <$ 2.4  &     0.123 $\pm$    0.016&     1.526 $\pm$    0.271&     0.483 $\pm$    0.080&     0.043 $\pm$    0.001 &     3.022 $\pm$    0.974&      13.25 / 152 &     1.00  \\ \hline 
-3.0 $< \eta <$ 3.0  &     0.034 $\pm$    0.001&     1.883 $\pm$    0.632&     0.664 $\pm$    0.083&     0.129 $\pm$    0.019 &     2.543 $\pm$    0.780&      19.43 / 183 &     1.00  \\ \hline 
-3.4 $< \eta <$ 3.4  &     0.031 $\pm$    0.001&     2.122 $\pm$    0.665&     0.638 $\pm$    0.078&     0.112 $\pm$    0.015 &     2.506 $\pm$    0.672&      21.18 / 204 &     1.00  \\ \hline 
-3.4 $< \eta <$ 5.0  &     0.027 $\pm$    0.001&     2.143 $\pm$    0.799&     0.663 $\pm$    0.093&     0.101 $\pm$    0.016 &     3.206 $\pm$    1.109&      23.36 / 231 &     1.00  \\ \hline 
\hline 
 \multicolumn{8}{|c|}{INEL$>$0  (Forward)}\\ 
 \hline-2.0 $< \eta <$ 2.0  &     0.145 $\pm$    0.009&     1.807 $\pm$    0.145&     0.500 $\pm$    0.042&     0.052 $\pm$    0.001 &     3.484 $\pm$    0.595&      28.45 / 132 &     1.00  \\ \hline 
-2.4 $< \eta <$ 2.4  &     0.136 $\pm$    0.015&     2.213 $\pm$    0.381&     0.414 $\pm$    0.076&     0.043 $\pm$    0.001 &     2.623 $\pm$    0.788&      14.73 / 152 &     1.00  \\ \hline 
-3.0 $< \eta <$ 3.0  &     0.034 $\pm$    0.001&     1.715 $\pm$    0.603&     0.722 $\pm$    0.087&     0.138 $\pm$    0.023 &     3.645 $\pm$    1.369&      20.48 / 183 &     1.00  \\ \hline 
-3.4 $< \eta <$ 3.4  &     0.031 $\pm$    0.001&     1.975 $\pm$    0.643&     0.689 $\pm$    0.082&     0.118 $\pm$    0.017 &     3.384 $\pm$    1.031&      22.31 / 204 &     1.00  \\ \hline 
-3.4 $< \eta <$ 5.0  &     0.027 $\pm$    0.001&     1.927 $\pm$    0.792&     0.724 $\pm$    0.102&     0.108 $\pm$    0.020 &     4.457 $\pm$    2.009&      21.38 / 230 &     1.00  \\ \hline 
\hline 
 \multicolumn{8}{|c|}{NSD  (Forward)}\\ 
 \hline-2.0 $< \eta <$ 2.0  &     0.125 $\pm$    0.009&     1.737 $\pm$    0.103&     0.518 $\pm$    0.048&     0.050 $\pm$    0.001 &     3.973 $\pm$    0.759&      22.15 / 132 &     1.00  \\ \hline 
-2.4 $< \eta <$ 2.4  &     0.123 $\pm$    0.014&     2.163 $\pm$    0.310&     0.409 $\pm$    0.089&     0.041 $\pm$    0.001 &     2.768 $\pm$    0.935&      13.21 / 153 &     1.00  \\ \hline 
-3.0 $< \eta <$ 3.0  &     0.033 $\pm$    0.001&     1.573 $\pm$    0.478&     0.774 $\pm$    0.079&     0.138 $\pm$    0.024 &     4.280 $\pm$    1.815&      18.40 / 184 &     1.00  \\ \hline 
-3.4 $< \eta <$ 3.4  &     0.030 $\pm$    0.001&     1.581 $\pm$    0.455&     0.777 $\pm$    0.072&     0.126 $\pm$    0.020 &     4.756 $\pm$    1.943&      20.92 / 205 &     1.00  \\ \hline 
-3.4 $< \eta <$ 5.0  &     0.026 $\pm$    0.000&     1.460 $\pm$    0.211&     0.831 $\pm$    0.033&     0.130 $\pm$    0.014 &     8.809 $\pm$    2.778&      23.00 / 232 &     1.00  \\ \hline 
\end{tabular}}
}
\end{center}
 \end{table} 
 
 \begin{table}[pt]
 \begin{center}
 \scalebox{0.8}{ 
 \tbl{Fit parameters, $\chi^{2}/ndf$ and $p{\text -}$values obtained from 2SGD for 7~TeV inelastic and single${\text -}$non${\text -}$diffractive events.}  
 {\begin{tabular}{@{}|c c c c c c c c|@{}} \toprule
$\eta$ &   $b_{1}$ & $t_{1}$&   $\alpha$ & $b_{2}$&   $t_{2}$   & $\chi^2/ndf$ & $p$-$value$  \\ 
 \hline 
\hline 
 \multicolumn{8}{|c|}{INEL}\\ 
 \hline-0.5 $< \eta <$ 0.5  &     0.223 $\pm$    0.005&     0.378 $\pm$    0.044&     0.935 $\pm$    0.009&     0.222 $\pm$    0.003 &    30.162 $\pm$    4.910&      62.49 / 61 &     0.42  \\ \hline 
-1.0 $< \eta <$ 1.0  &     0.122 $\pm$    0.002&     0.476 $\pm$    0.034&     0.913 $\pm$    0.005&     0.121 $\pm$    0.001 &    27.457 $\pm$    1.837&     260.47 / 109 &      $<$0.01  \\ \hline 
-1.5 $< \eta <$ 1.5  &     0.085 $\pm$    0.002&     0.473 $\pm$    0.043&     0.895 $\pm$    0.009&     0.079 $\pm$    0.001 &    18.612 $\pm$    1.388&     197.35 / 146 &     $<$0.01  \\ \hline 
\hline 
 \multicolumn{8}{|c|}{INEL$>$0}\\ 
 \hline-1.0 $< \eta <$ 1.0  &     0.123 $\pm$    0.002&     0.471 $\pm$    0.034&     0.911 $\pm$    0.005&     0.121 $\pm$    0.001 &    26.900 $\pm$    1.804&     255.80 / 109 &    $<$0.01 \\ \hline 
\hline 
 \multicolumn{8}{|c|}{NSD}\\ 
 \hline-0.5 $< \eta <$ 0.5  &     0.226 $\pm$    0.008&     0.702 $\pm$    0.178&     0.950 $\pm$    0.013&     0.227 $\pm$    0.006 &    50.604 $\pm$   17.312&      23.65 / 59 &     1.00  \\ \hline 
-1.0 $< \eta <$ 1.0  &     0.125 $\pm$    0.003&     0.732 $\pm$    0.051&     0.924 $\pm$    0.009&     0.124 $\pm$    0.002 &    42.052 $\pm$    6.068&      92.05 / 109 &     0.88  \\ \hline 
-1.5 $< \eta <$ 1.5  &     0.090 $\pm$    0.005&     0.862 $\pm$    0.086&     0.895 $\pm$    0.019&     0.081 $\pm$    0.001 &    25.185 $\pm$    3.890&      64.06 / 145 &     1.00  \\ \hline 
\hline  
 \multicolumn{8}{|c|}{INEL (Forward)}\\ 
 \hline-2.0 $< \eta <$ 2.0  &     0.051 $\pm$    0.001&     3.061 $\pm$    0.593&     0.515 $\pm$    0.047&     0.146 $\pm$    0.010 &     1.629 $\pm$    0.148&      17.99 / 133 &     1.00  \\ \hline 
-2.4 $< \eta <$ 2.4  &     0.044 $\pm$    0.001&     2.761 $\pm$    0.740&     0.558 $\pm$    0.069&     0.134 $\pm$    0.014 &     1.968 $\pm$    0.337&      18.59 / 158 &     1.00  \\ \hline 
-3.0 $< \eta <$ 3.0  &     0.036 $\pm$    0.001&     2.687 $\pm$    0.836&     0.592 $\pm$    0.084&     0.117 $\pm$    0.017 &     2.361 $\pm$    0.580&      26.55 / 192 &     1.00  \\ \hline 
-3.4 $< \eta <$ 3.4  &     0.033 $\pm$    0.001&     2.595 $\pm$    0.858&     0.607 $\pm$    0.088&     0.108 $\pm$    0.016 &     2.590 $\pm$    0.699&      23.70 / 208 &     1.00  \\ \hline 
-3.4 $< \eta <$ 5.0  &     0.029 $\pm$    0.001&     2.885 $\pm$    0.950&     0.611 $\pm$    0.088&     0.096 $\pm$    0.015 &     3.056 $\pm$    0.866&      28.34 / 238 &     1.00  \\ \hline 
\hline 
 \multicolumn{8}{|c|}{INEL$>$0  (Forward)}\\ 
 \hline-2.0 $< \eta <$ 2.0  &     0.051 $\pm$    0.001&     2.771 $\pm$    0.517&     0.576 $\pm$    0.047&     0.157 $\pm$    0.010 &     2.320 $\pm$    0.202&      18.42 / 133 &     1.00  \\ \hline 
-2.4 $< \eta <$ 2.4  &     0.141 $\pm$    0.015&     2.688 $\pm$    0.489&     0.385 $\pm$    0.074&     0.043 $\pm$    0.001 &     2.552 $\pm$    0.702&      19.02 / 158 &     1.00  \\ \hline 
-3.0 $< \eta <$ 3.0  &     0.033 $\pm$    0.001&     1.090 $\pm$    0.202&     0.843 $\pm$    0.038&     0.181 $\pm$    0.025 &     8.403 $\pm$    3.481&      15.89 / 169 &     1.00  \\ \hline 
-3.4 $< \eta <$ 3.4  &     0.033 $\pm$    0.001&     2.530 $\pm$    1.087&     0.647 $\pm$    0.118&     0.109 $\pm$    0.021 &     3.176 $\pm$    1.117&      27.42 / 210 &     1.00  \\ \hline 
-3.4 $< \eta <$ 5.0  &     0.027 $\pm$    0.001&     1.301 $\pm$    0.245&     0.846 $\pm$    0.040&     0.150 $\pm$    0.023 &    11.325 $\pm$    5.584&      27.94 / 233 &     1.00  \\ \hline 
\hline 
 \multicolumn{8}{|c|}{NSD  (Forward)}\\ 
 \hline-2.0 $< \eta <$ 2.0  &     0.159 $\pm$    0.010&     2.225 $\pm$    0.186&     0.410 $\pm$    0.045&     0.051 $\pm$    0.001 &     2.625 $\pm$    0.480&      19.61 / 133 &     1.00  \\ \hline 
-2.4 $< \eta <$ 2.4  &     0.143 $\pm$    0.015&     2.703 $\pm$    0.477&     0.375 $\pm$    0.069&     0.043 $\pm$    0.001 &     2.448 $\pm$    0.637&      18.64 / 158 &     1.00  \\ \hline 
-3.0 $< \eta <$ 3.0  &     0.035 $\pm$    0.001&     2.322 $\pm$    0.695&     0.667 $\pm$    0.080&     0.128 $\pm$    0.018 &     3.416 $\pm$    0.900&      23.04 / 192 &     1.00  \\ \hline 
-3.4 $< \eta <$ 3.4  &     0.032 $\pm$    0.001&     2.184 $\pm$    0.820&     0.686 $\pm$    0.099&     0.119 $\pm$    0.021 &     3.773 $\pm$    1.355&      22.27 / 209 &     1.00  \\ \hline 
-3.4 $< \eta <$ 5.0  &     0.028 $\pm$    0.001&     2.233 $\pm$    0.955&     0.718 $\pm$    0.111&     0.112 $\pm$    0.025 &     5.190 $\pm$    2.618&      26.53 / 240 &     1.00  \\ \hline 
\end{tabular}}
} 
\end{center}
 \end{table}

 \begin{table}[pt]  
 \begin{center}
 \scalebox{0.8}{
 \tbl{Fit parameter values, $\chi^{2}/ndf$ and $p{\text -}$values obtained from 2SGD for 2.76~TeV inelastic and single$p{\text -}$non${\text -}$diffractive events.}  
 {\begin{tabular}{@{}|c c c c c c c c|@{}} \toprule 
$\eta$ &   $b_{1}$ & $t_{1}$&   $\alpha$ & $b_{2}$&   $t_{2}$   & $\chi^2/ndf$ & $p$-$value$  \\ 
 \hline 
\hline 
 \multicolumn{8}{|c|}{INEL}\\ 
 \hline-0.5 $< \eta <$ 0.5  &     0.280 $\pm$    0.013&     0.508 $\pm$    0.076&     0.945 $\pm$    0.014&     0.271 $\pm$    0.008 &    26.075 $\pm$    6.921&      12.30 / 42&       1.00   \\ \hline 
-1.0 $< \eta <$ 1.0  &     0.164 $\pm$    0.008&     0.729 $\pm$    0.080&     0.910 $\pm$    0.015&     0.149 $\pm$    0.002 &    23.070 $\pm$    3.131&      43.41 / 76&       1.00   \\ \hline 
-1.5 $< \eta <$ 1.5  &     0.158 $\pm$    0.008&     1.374 $\pm$    0.078&     0.783 $\pm$    0.025&     0.102 $\pm$    0.001 &    11.586 $\pm$    1.509&      54.32 / 98&       1.00   \\ \hline 
\hline 
 \multicolumn{8}{|c|}{INEL $> $0}\\ 
 \hline-1.0 $< \eta <$ 1.0  &     0.165 $\pm$    0.008&     0.748 $\pm$    0.078&     0.909 $\pm$    0.015&     0.149 $\pm$    0.002 &    22.764 $\pm$    3.118&      45.90 / 76&       1.00   \\ \hline 
\hline 
 \multicolumn{8}{|c|}{NSD}\\ 
 \hline-0.5 $< \eta <$ 0.5  &     0.284 $\pm$    0.024&     0.691 $\pm$    0.127&     0.948 $\pm$    0.028&     0.274 $\pm$    0.014 &    32.830 $\pm$   17.546&       4.10 / 43&       1.00   \\ \hline 
-1.0 $< \eta <$ 1.0  &     0.178 $\pm$    0.019&     1.145 $\pm$    0.179&     0.897 $\pm$    0.038&     0.151 $\pm$    0.002 &    26.680 $\pm$    7.922&      15.93 / 76&       1.00   \\ \hline 
-1.5 $< \eta <$ 1.5  &     0.134 $\pm$    0.013&     1.440 $\pm$    0.165&     0.857 $\pm$    0.043&     0.103 $\pm$    0.002 &    19.700 $\pm$    5.433&      22.97 / 98&       1.00   \\ \hline 
\end{tabular}} 
}
\end{center}
 \end{table} 
 \subsection{Moments of multiplicity distributions} 
\begin{figure}[th]
\centerline{\includegraphics[scale=0.4]{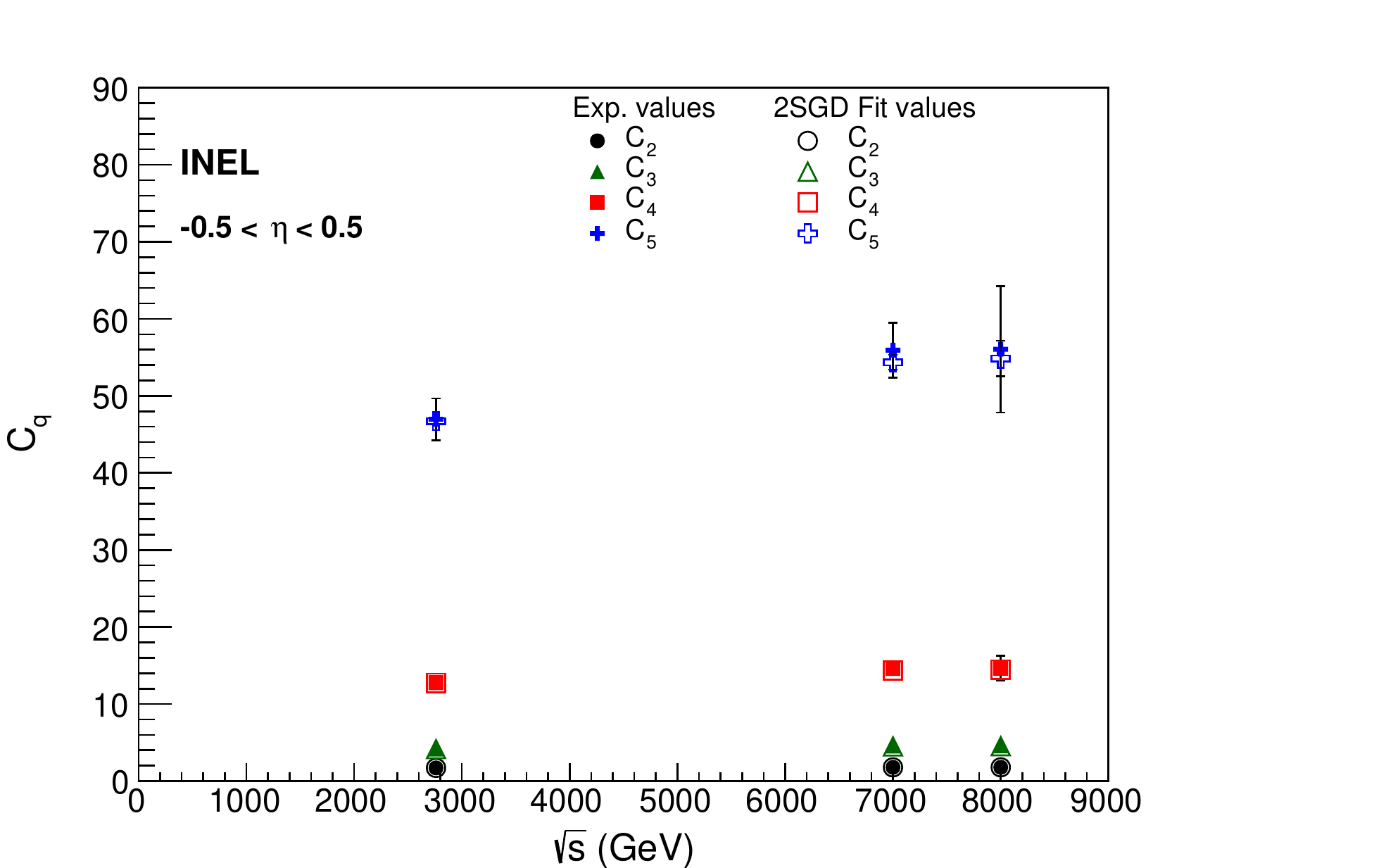}\hspace{-1.0cm}\includegraphics[scale=0.4]{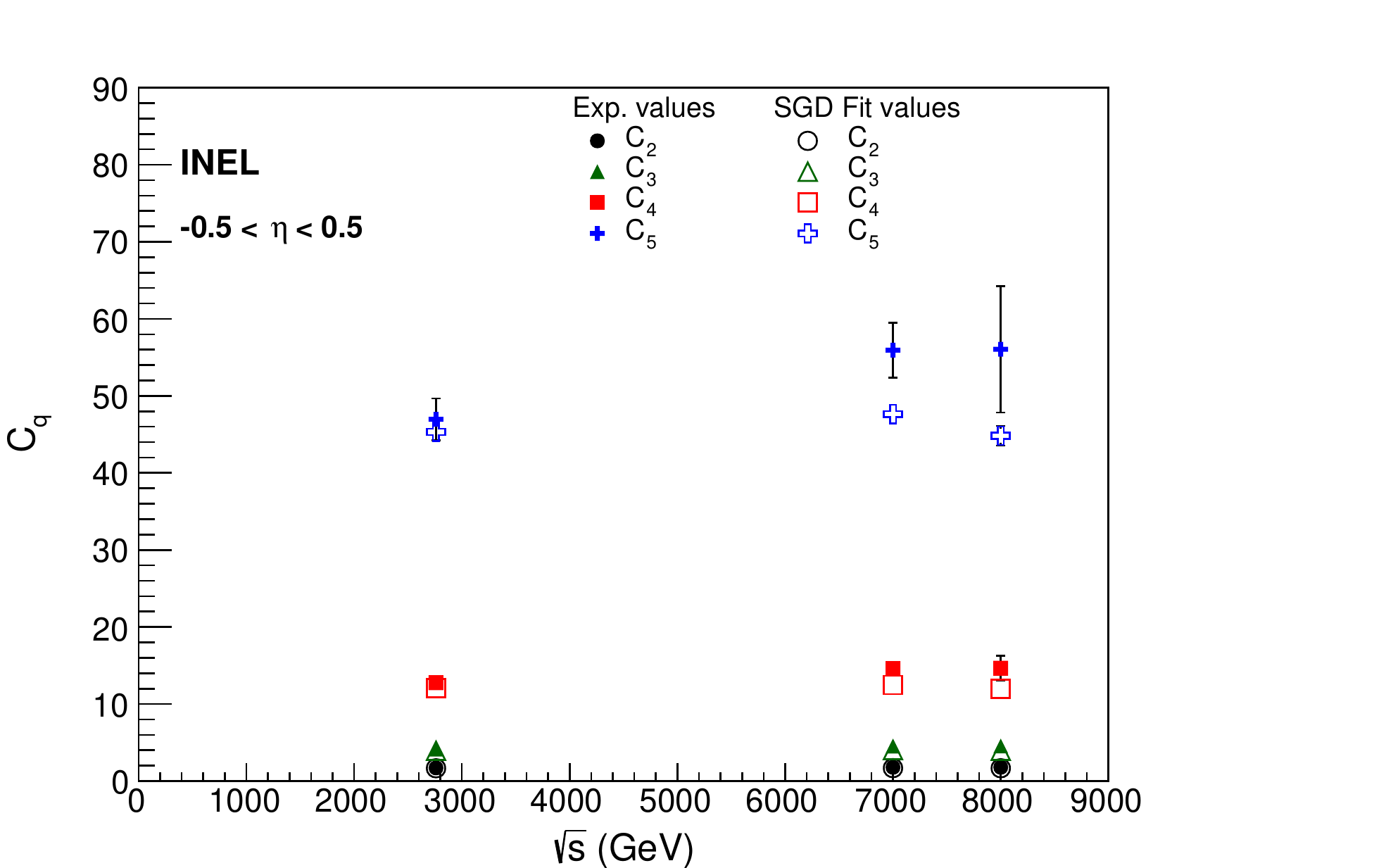}}
\centerline{\includegraphics[scale=0.4]{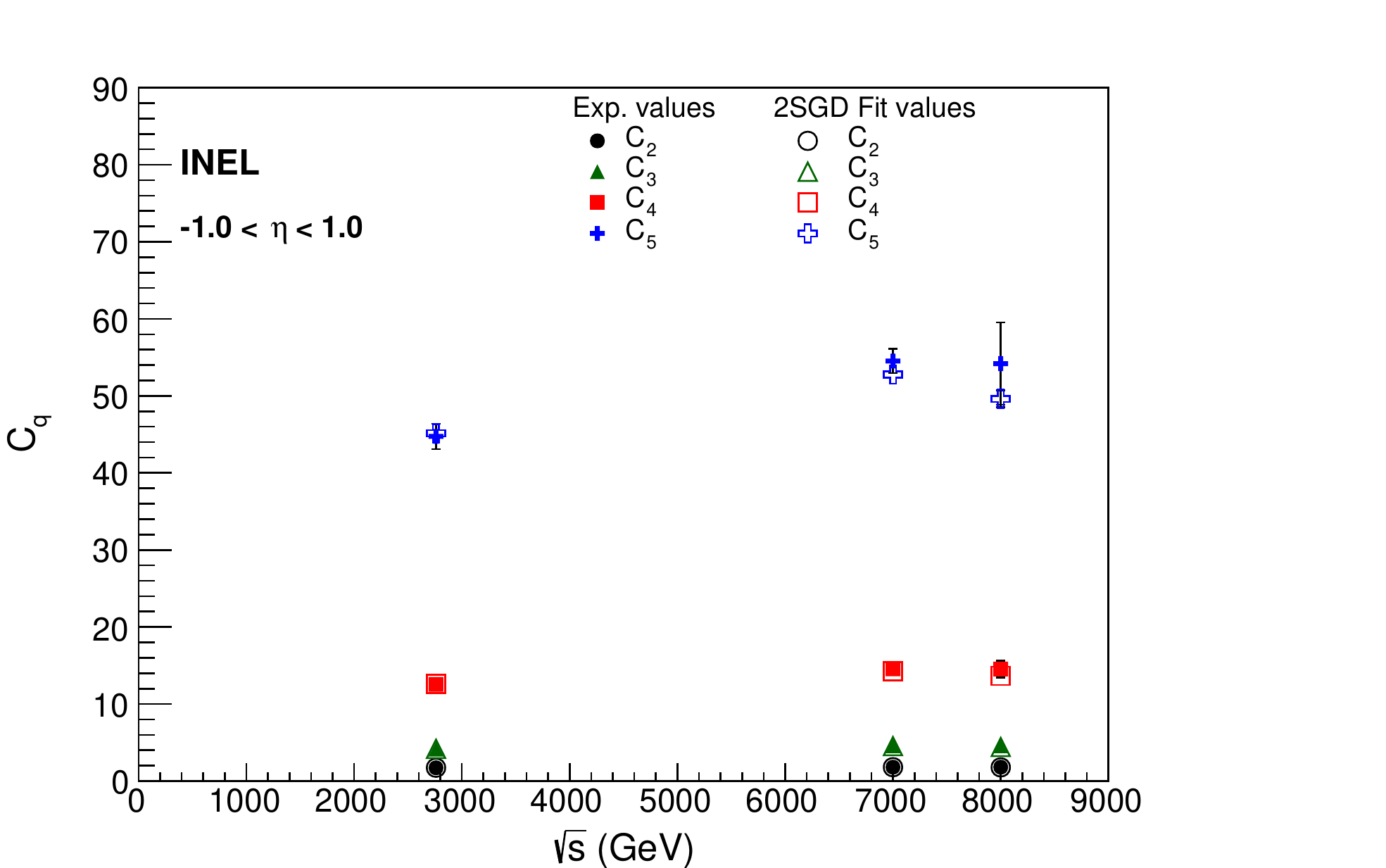}\hspace{-1.0cm}\includegraphics[scale=0.4]{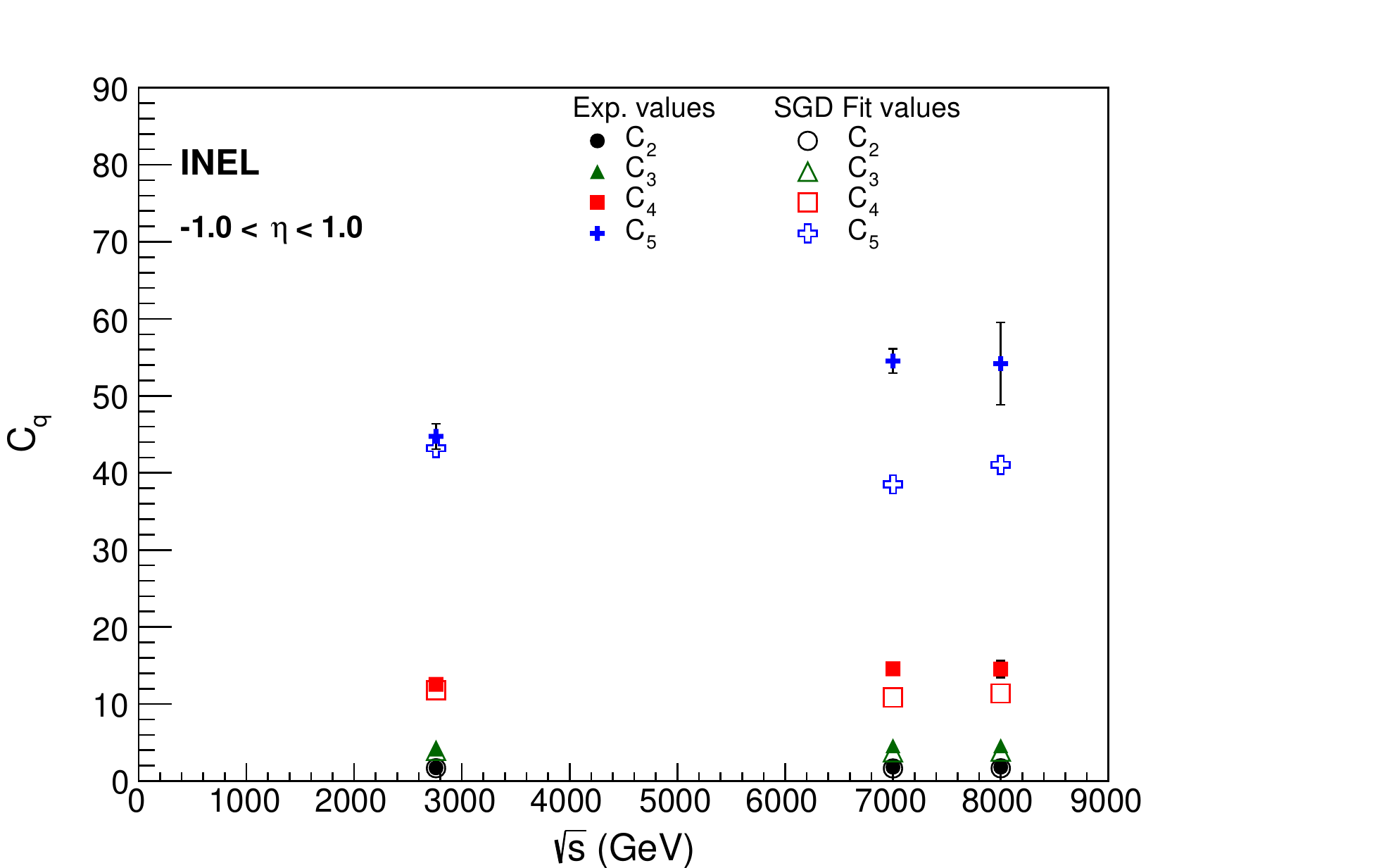}}
\centerline{\includegraphics[scale=0.4]{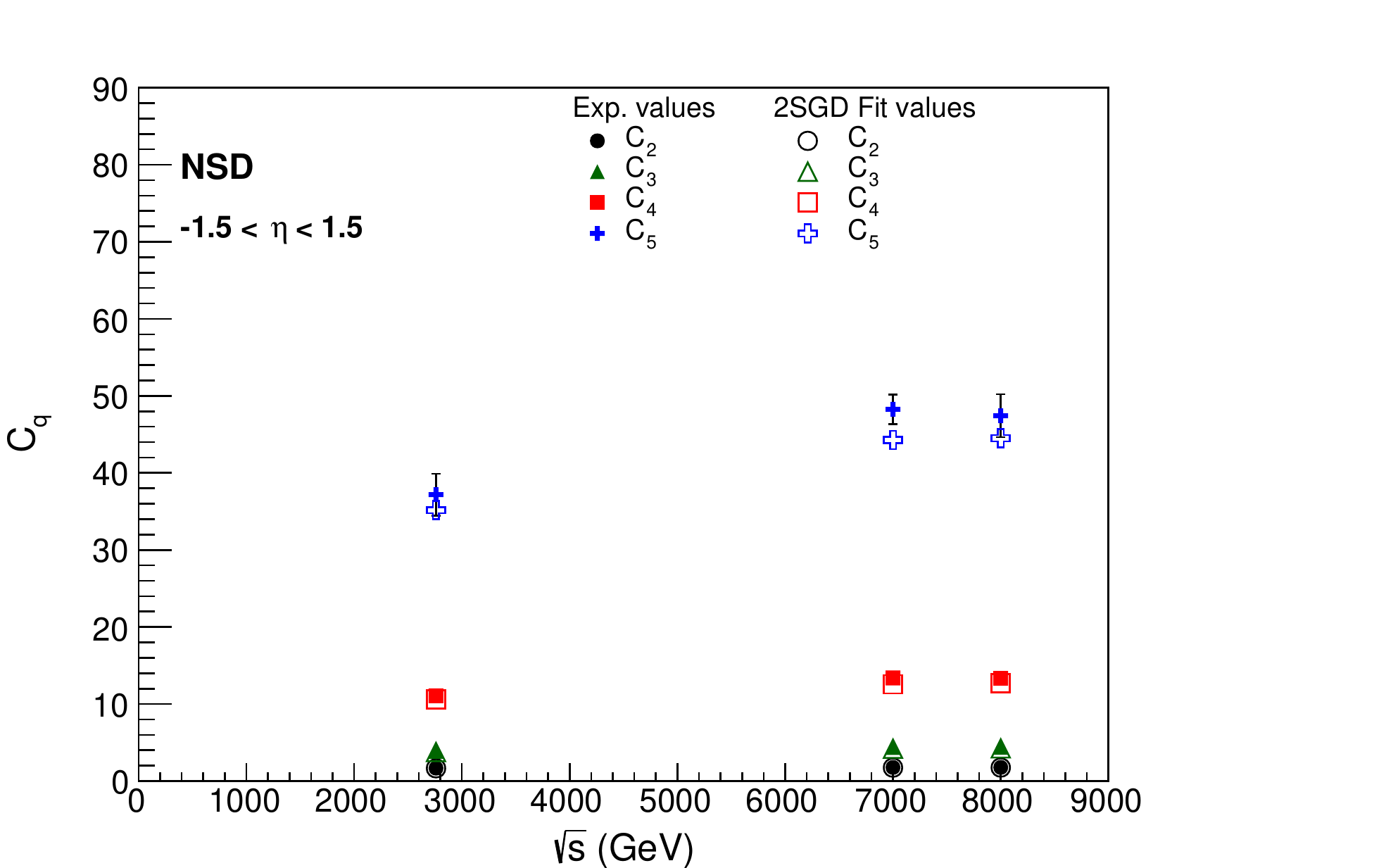}\hspace{-1.0cm}\includegraphics[scale=0.4]{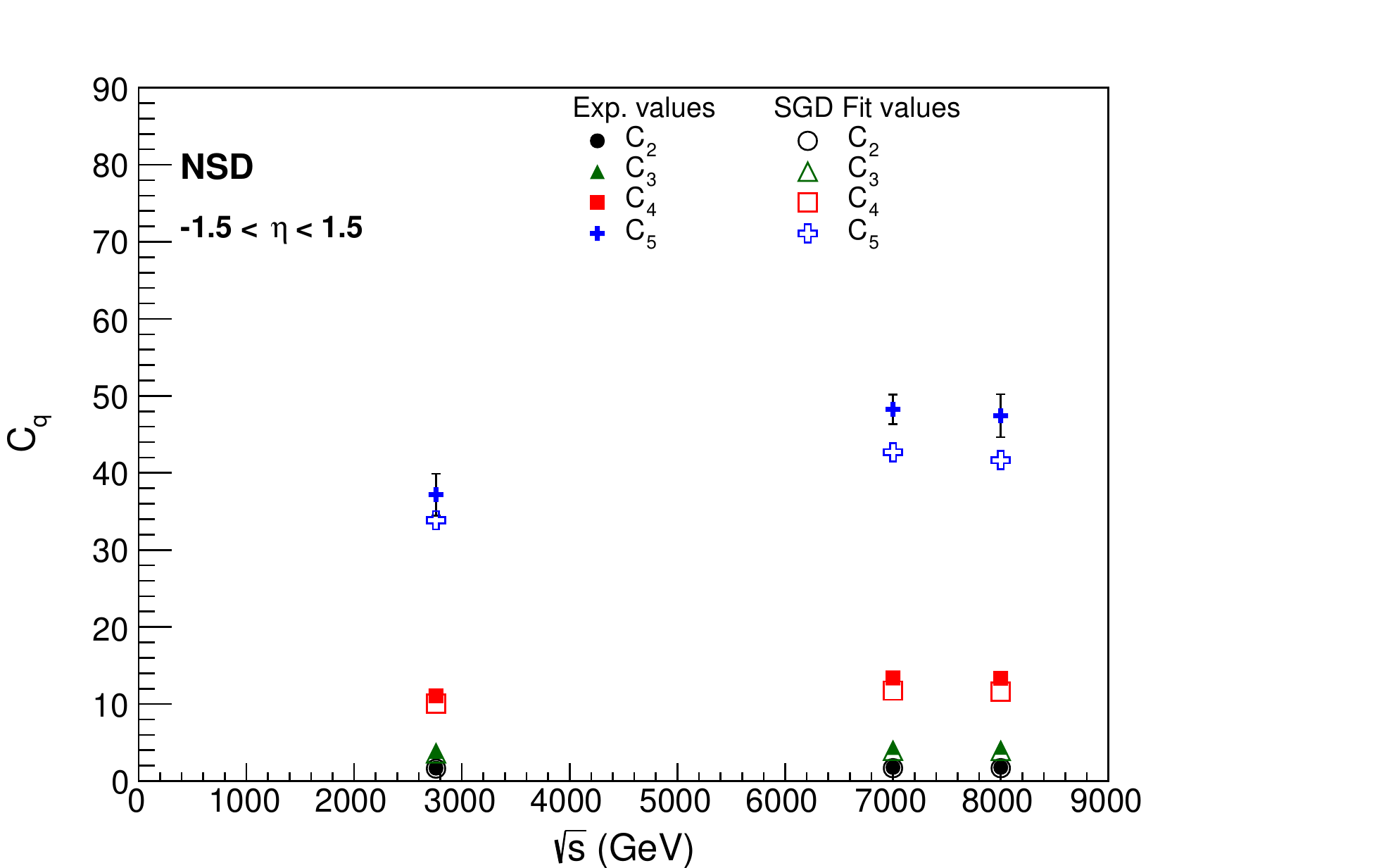}}
\caption{Normalised moments of $n{\text -}$charged multiplicity in 8,7 and 2.76~TeV inelastic events in the pseudorapidity $|\eta|<$ 0.5 and $|\eta|<$ 1.0 windows and for the NSD events in the pseudorapidity $|\eta|<$1.5 region of the ALICE detector at the LHC.~Data are compared to the 2SGD\,(left) and SGD\,(right) predictions.}
\label{figCq}
\end{figure}

Study of the moments of multiplicity distribution allows to analyse the shape and scale of a distribution.~Moments-analysis also serves as an important tool to study these properties as a function of the center${\text -}$of${\text -}$mass~(c.m.s) energy and to check the validity of KNO scaling.~Figure~\ref{figCq} shows the normalized moments $C_{q}$ for $q$\,=\,2,\,3,\,4,\,5 for $\sqrt{s}$ = 8,\,7 and 2.76~TeV for inelastic and NSD collisions in three pseudorapidity intervals, $|\eta|<$ 0.5, $|\eta|<$ 1.0 and $|\eta|<$ 1.5.~The values obtained by fitting the 2SGD and SGD are shown in comparison to the experimental values obtained from the probability distributions.~It may be observed from the figures and the values in the tables~5-8 that $C_2$ remains constant over the energy range, $C_3$ shows a small increase with increasing energy for the two largest $\eta$ windows, $C_4$ and $C_5$ show an increase with increasing energy, which becomes stronger for larger $\eta$ windows.
~However,the values calculated from 2SGD fits reproduce the experimental values very precisely. 

\begin{figure}[th]
\centerline{\includegraphics[scale=0.4]{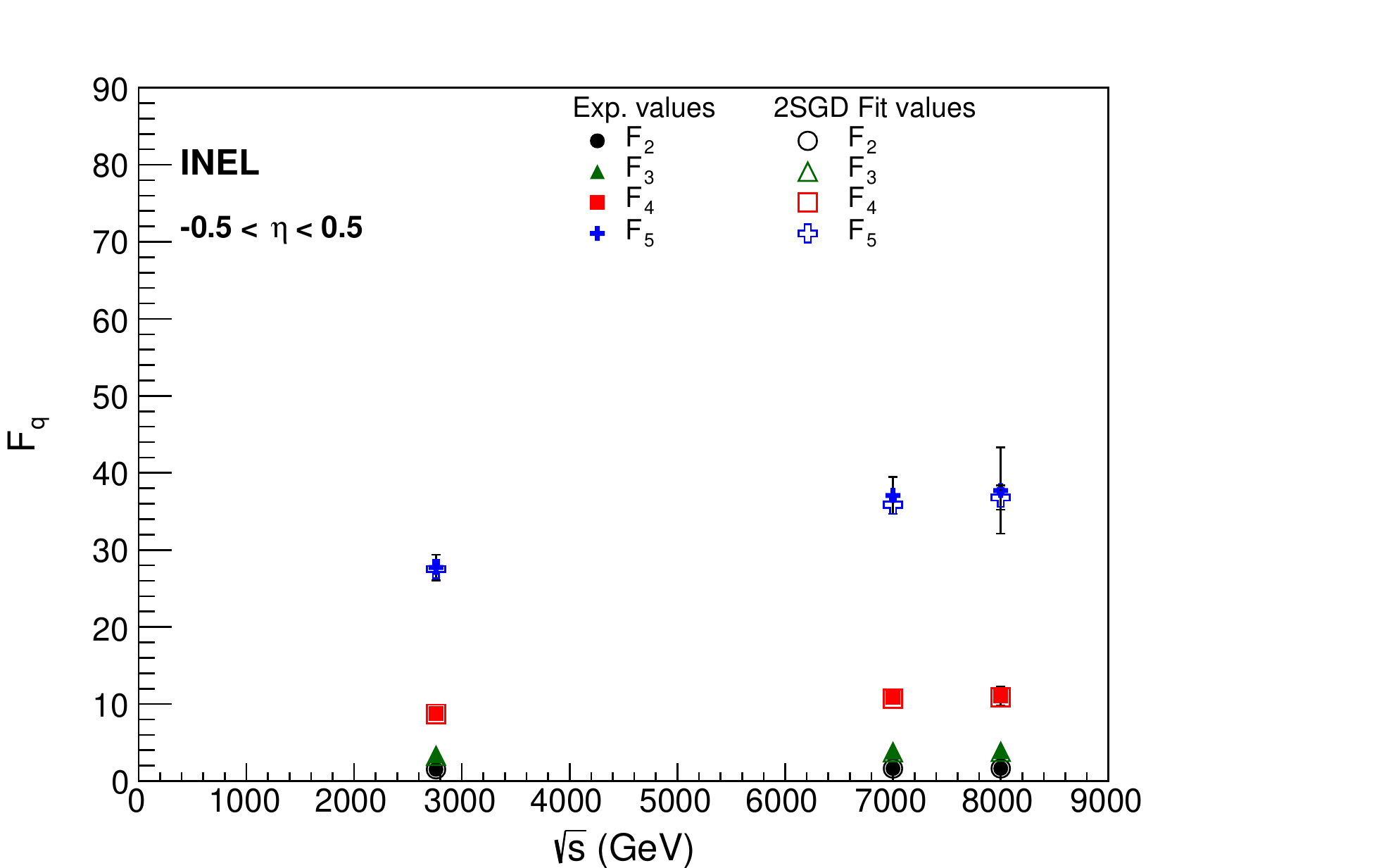}\hspace{-1.0cm}\includegraphics[scale=0.4]{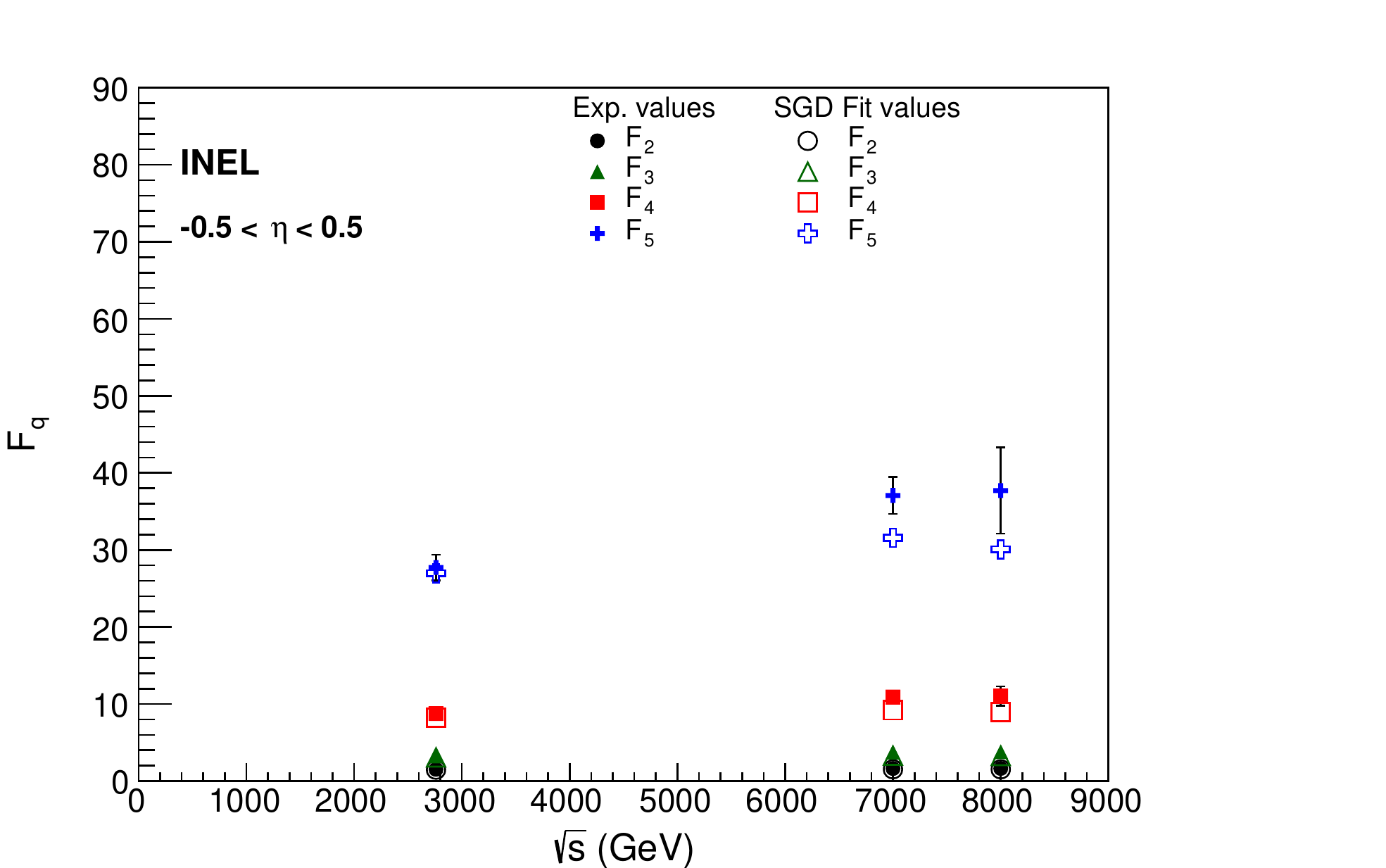}}
\centerline{\includegraphics[scale=0.4]{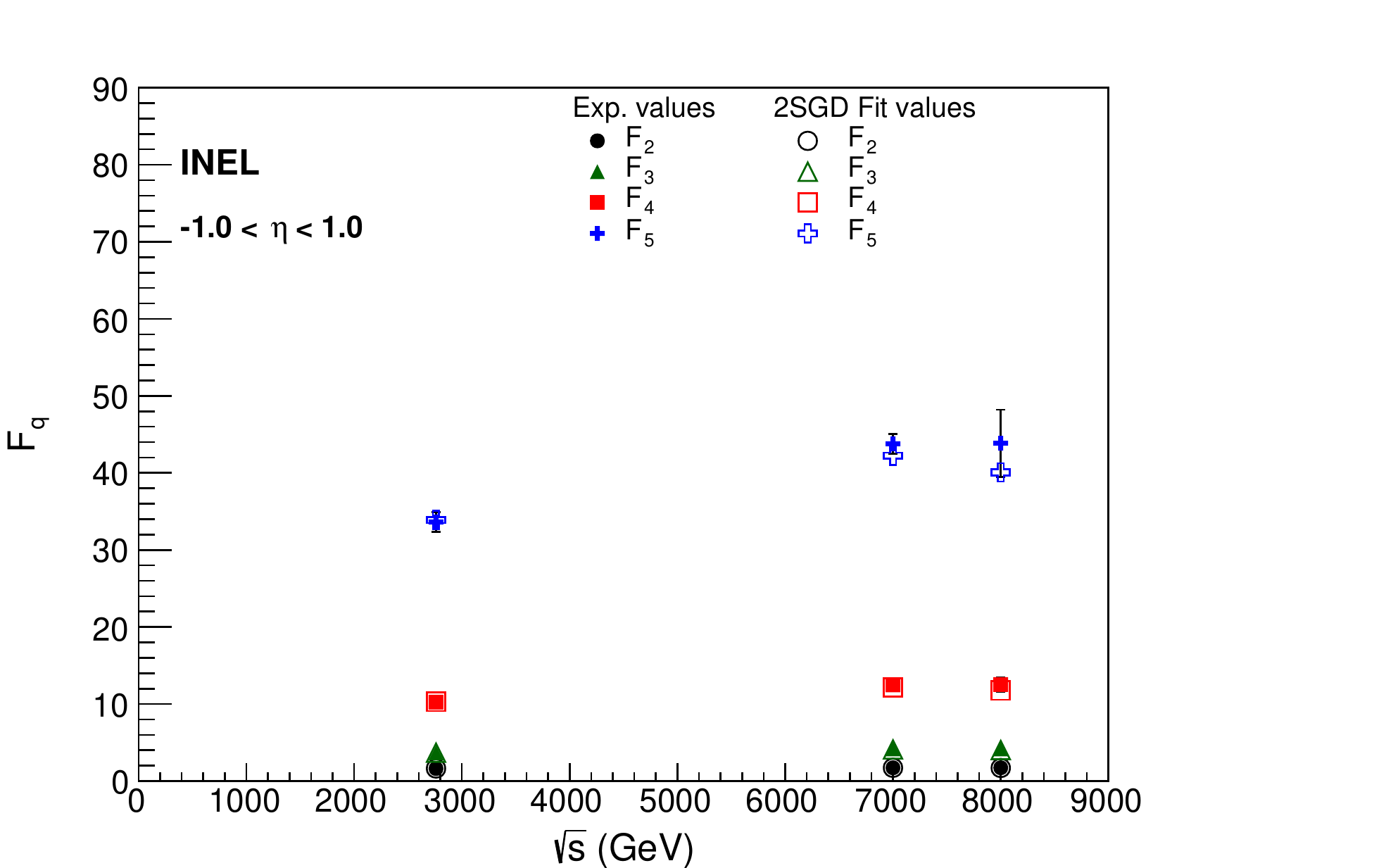}\hspace{-1.0cm}\includegraphics[scale=0.4]{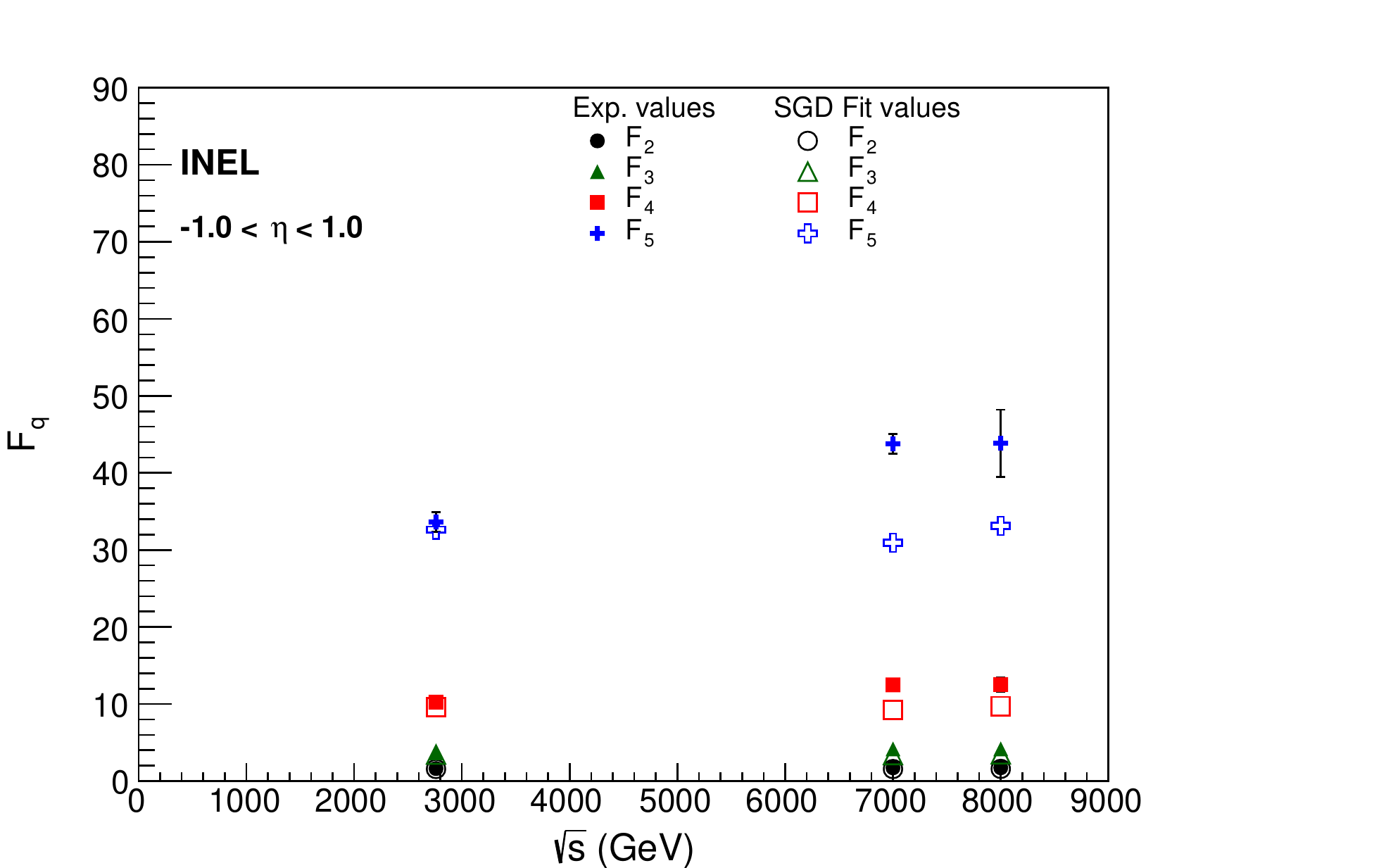}}
\centerline{\includegraphics[scale=0.4]{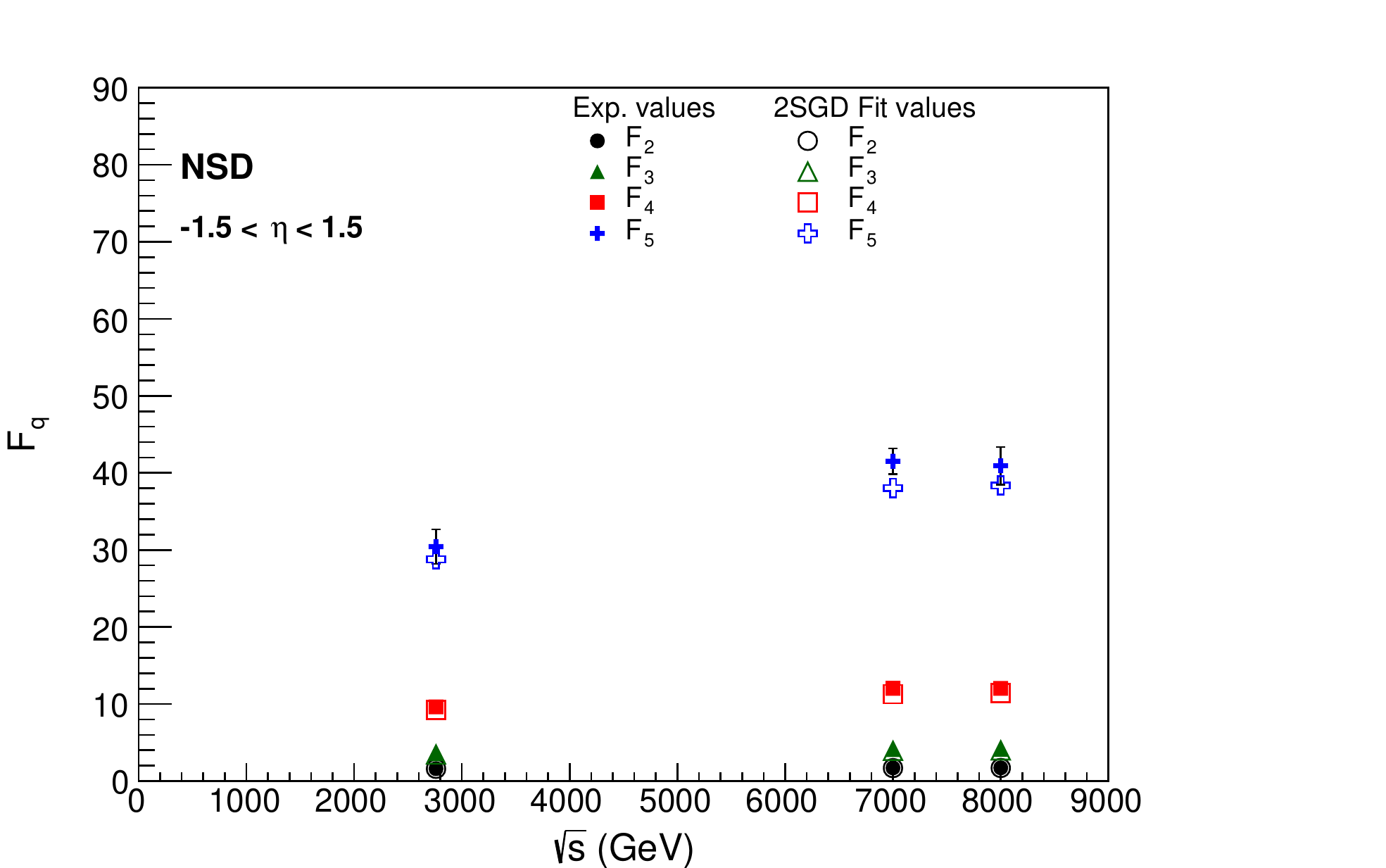}\hspace{-1.0cm}\includegraphics[scale=0.4]{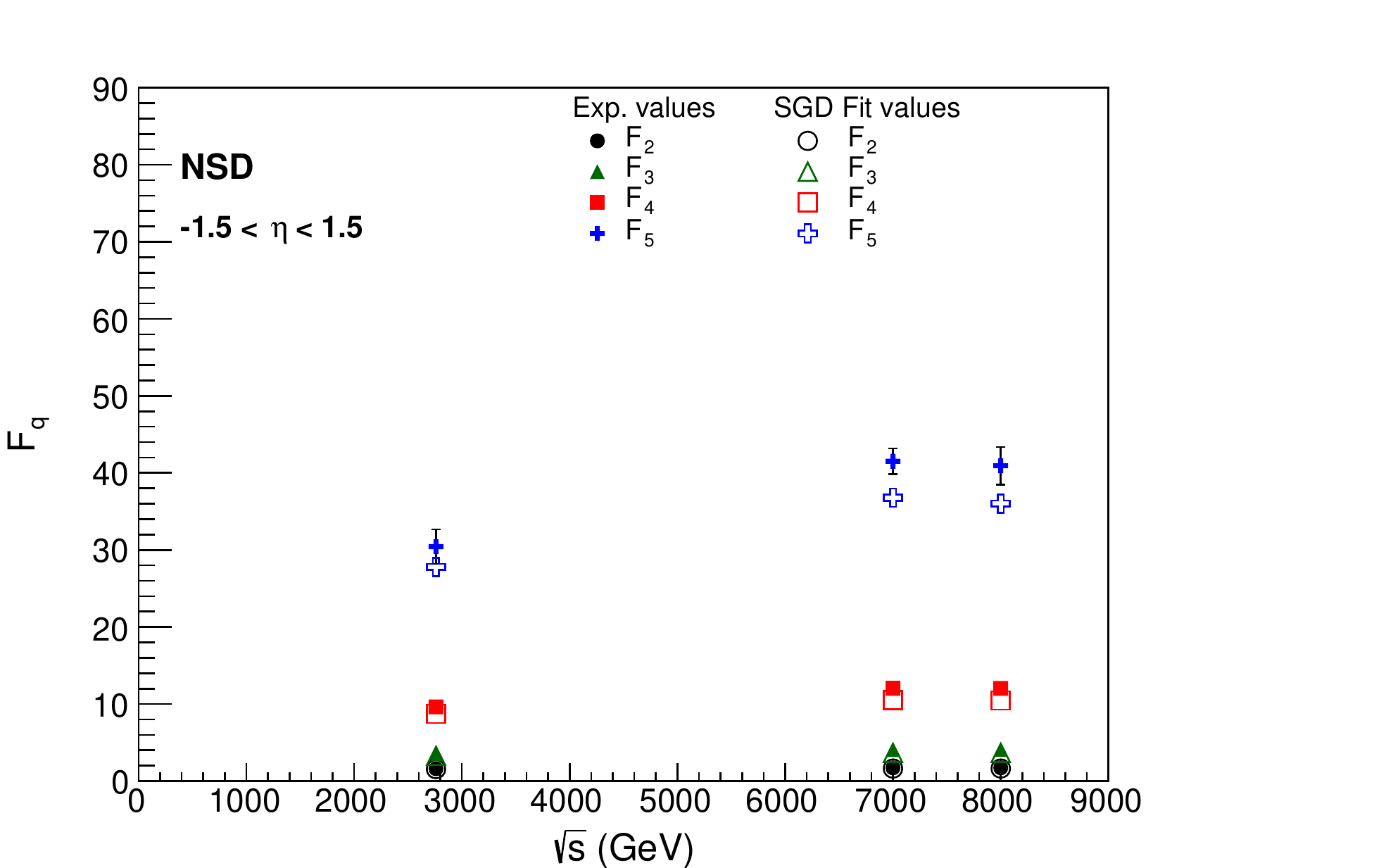}}
\caption{Normalised factorial moments of $n{\text -}$charged multiplicity in 8,7 and 2.76~TeV inelastic events in the pseudorapidity $|\eta|<$ 0.5 and $|\eta|<$ 1.0 windows and for the NSD events in the pseudorapidity $|\eta|<$1.5 region of the ALICE detector at the LHC.~Data are compared to the 2SGD\,(left) and SGD\,(right) predictions.}
\label{figFq}
\end{figure} 
 \begin{table}[pt]
 
 \begin{center}
 \scalebox{0.8}{ 
 \tbl{Normalized moments and normalized factorial moments of multiplicity distributions of
 inelastic and non-single-diffractive events obtained from the data at $\sqrt{s}$=8~TeV.~The uncertainties are combined statistical and systematic uncertainties.} 
 {\begin{tabular}{@{}|c c c c c c c c c| @{}}  \toprule 
$\eta$ &  $C_2$ & $C_3$  & $C_4$ & $C_5$ &  $F_2$ & $F_3$  & $F_4$ & $F_5$ \\ 
 \hline 
\hline 
 \multicolumn{9}{|c|}{INEL}\\ 
 \hline-0.5 $< \eta <$ 0.5 &      1.788 $\pm$    0.066&       4.55 $\pm$     0.34&      14.65 $\pm$     1.62&      56.05 $\pm$     8.20&      1.641 $\pm$    0.062&       3.81 $\pm$     0.29 &     11.04 $\pm$     1.24 &     37.70 $\pm$     5.60 \\ \hline 
-1.0 $< \eta <$ 1.0 &      1.806 $\pm$    0.047&       4.59 $\pm$     0.23&      14.55 $\pm$     1.09&      54.19 $\pm$     5.34&      1.729 $\pm$    0.045&       4.18 $\pm$     0.21 &     12.53 $\pm$     0.95 &     43.85 $\pm$     4.35 \\ \hline 
-1.5 $< \eta <$ 1.5 &      1.807 $\pm$    0.036&       4.56 $\pm$     0.18&      14.26 $\pm$     0.83&      51.97 $\pm$     4.01&      1.754 $\pm$    0.035&       4.28 $\pm$     0.17 &     12.87 $\pm$     0.76 &     44.85 $\pm$     3.48 \\ \hline 
\hline 
 \multicolumn{9}{|c|}{INEL$>$0}\\ 
 \hline-1.0 $< \eta <$ 1.0 &      1.807 $\pm$    0.007&       4.59 $\pm$     0.04&      14.54 $\pm$     0.19&      54.18 $\pm$     0.93&      1.729 $\pm$    0.007&       4.18 $\pm$     0.04 &     12.53 $\pm$     0.16 &     43.83 $\pm$     0.76 \\ \hline 
\hline 
 \multicolumn{9}{|c|}{NSD}\\ 
 \hline-0.5 $< \eta <$ 0.5 &      1.771 $\pm$    0.048&       4.45 $\pm$     0.24&      14.13 $\pm$     1.16&      53.35 $\pm$     5.84&      1.626 $\pm$    0.045&       3.72 $\pm$     0.21 &     10.65 $\pm$     0.89 &     35.89 $\pm$     3.99 \\ \hline 
-1.0 $< \eta <$ 1.0 &      1.778 $\pm$    0.035&       4.43 $\pm$     0.17&      13.75 $\pm$     0.79&      50.19 $\pm$     3.78&      1.703 $\pm$    0.034&       4.04 $\pm$     0.16 &     11.85 $\pm$     0.68 &     40.61 $\pm$     3.08 \\ \hline 
-1.5 $< \eta <$ 1.5 &      1.772 $\pm$    0.027&       4.37 $\pm$     0.13&      13.34 $\pm$     0.60&      47.42 $\pm$     2.80&      1.721 $\pm$    0.026&       4.10 $\pm$     0.12 &     12.03 $\pm$     0.54 &     40.93 $\pm$     2.43 \\ \hline 
\hline  
 \multicolumn{9}{|c|}{INEL (Forward)}\\ 
 \hline-2.0 $< \eta <$ 2.0 &      1.811 $\pm$    0.016&       4.51 $\pm$     0.08&      13.64 $\pm$     0.39&      46.88 $\pm$     1.81&      1.771 $\pm$    0.015&       4.30 $\pm$     0.08 &     12.59 $\pm$     0.36 &     41.67 $\pm$     1.62 \\ \hline 
-2.4 $< \eta <$ 2.4 &      1.749 $\pm$    0.022&       4.18 $\pm$     0.11&      12.07 $\pm$     0.53&      39.45 $\pm$     2.37&      1.717 $\pm$    0.022&       4.01 $\pm$     0.11 &     11.27 $\pm$     0.49 &     35.69 $\pm$     2.16 \\ \hline 
-3.0 $< \eta <$ 3.0 &      1.709 $\pm$    0.023&       3.97 $\pm$     0.11&      11.08 $\pm$     0.51&      34.96 $\pm$     2.25&      1.684 $\pm$    0.022&       3.84 $\pm$     0.11 &     10.48 $\pm$     0.49 &     32.18 $\pm$     2.08 \\ \hline 
-3.4 $< \eta <$ 3.4 &      1.708 $\pm$    0.021&       3.96 $\pm$     0.11&      11.00 $\pm$     0.47&      34.57 $\pm$     2.10&      1.685 $\pm$    0.020&       3.84 $\pm$     0.10 &     10.46 $\pm$     0.45 &     32.08 $\pm$     1.96 \\ \hline 
-3.4 $< \eta <$ 5.0 &      1.689 $\pm$    0.022&       3.84 $\pm$     0.11&      10.41 $\pm$     0.50&      31.83 $\pm$     2.13&      1.669 $\pm$    0.022&       3.74 $\pm$     0.11 &      9.96 $\pm$     0.48 &     29.80 $\pm$     2.01 \\ \hline 
\hline 
 \multicolumn{9}{|c|}{INEL$>$0  (Forward)}\\ 
 \hline-2.0 $< \eta <$ 2.0 &      1.746 $\pm$    0.012&       4.17 $\pm$     0.06&      12.08 $\pm$     0.27&      39.79 $\pm$     1.22&      1.708 $\pm$    0.012&       3.97 $\pm$     0.06 &     11.15 $\pm$     0.25 &     35.37 $\pm$     1.09 \\ \hline 
-2.4 $< \eta <$ 2.4 &      1.703 $\pm$    0.017&       3.94 $\pm$     0.09&      10.99 $\pm$     0.38&      34.74 $\pm$     1.66&      1.671 $\pm$    0.017&       3.78 $\pm$     0.08 &     10.27 $\pm$     0.36 &     31.43 $\pm$     1.51 \\ \hline 
-3.0 $< \eta <$ 3.0 &      1.670 $\pm$    0.018&       3.77 $\pm$     0.09&      10.20 $\pm$     0.38&      31.21 $\pm$     1.62&      1.645 $\pm$    0.017&       3.64 $\pm$     0.08 &      9.64 $\pm$     0.36 &     28.73 $\pm$     1.50 \\ \hline 
-3.4 $< \eta <$ 3.4 &      1.667 $\pm$    0.016&       3.74 $\pm$     0.08&      10.09 $\pm$     0.36&      30.69 $\pm$     1.52&      1.644 $\pm$    0.016&       3.63 $\pm$     0.08 &      9.59 $\pm$     0.34 &     28.48 $\pm$     1.42 \\ \hline 
-3.4 $< \eta <$ 5.0 &      1.627 $\pm$    0.018&       3.53 $\pm$     0.09&       9.16 $\pm$     0.38&      26.71 $\pm$     1.54&      1.608 $\pm$    0.018&       3.44 $\pm$     0.09 &      8.76 $\pm$     0.36 &     25.01 $\pm$     1.45 \\ \hline 
\hline 
 \multicolumn{9}{|c|}{NSD  (Forward)}\\ 
 \hline2.4 $< \eta <$ 2.0 &      1.703 $\pm$    0.012&       3.92 $\pm$     0.06&      10.86 $\pm$     0.26&      34.04 $\pm$     1.13&      1.667 $\pm$    0.012&       3.74 $\pm$     0.06 &     10.04 $\pm$     0.25 &     30.34 $\pm$     1.01 \\ \hline 
-3.0 $< \eta <$ 2.4 &      1.684 $\pm$    0.018&       3.81 $\pm$     0.09&      10.36 $\pm$     0.38&      31.74 $\pm$     1.61&      1.654 $\pm$    0.018&       3.66 $\pm$     0.09 &      9.69 $\pm$     0.36 &     28.76 $\pm$     1.47 \\ \hline 
-3.4 $< \eta <$ 3.0 &      1.649 $\pm$    0.018&       3.64 $\pm$     0.09&       9.59 $\pm$     0.37&      28.46 $\pm$     1.55&      1.625 $\pm$    0.018&       3.52 $\pm$     0.09 &      9.08 $\pm$     0.36 &     26.23 $\pm$     1.43 \\ \hline 
-3.4 $< \eta <$ 3.4 &      1.643 $\pm$    0.017&       3.60 $\pm$     0.08&       9.44 $\pm$     0.35&      27.82 $\pm$     1.43&      1.621 $\pm$    0.016&       3.50 $\pm$     0.08 &      8.98 $\pm$     0.33 &     25.84 $\pm$     1.34 \\ \hline 
-3.4 $< \eta <$ 5.0 &      1.613 $\pm$    0.018&       3.44 $\pm$     0.09&       8.74 $\pm$     0.36&      24.88 $\pm$     1.44&      1.594 $\pm$    0.018&       3.35 $\pm$     0.09 &      8.36 $\pm$     0.35 &     23.32 $\pm$     1.36 \\ \hline 
\end{tabular}}
}
\end{center} 
 \end{table}
 \begin{table}[pt] 

 \begin{center}
 \scalebox{0.8}{
 \tbl{Normalized moments and normalized factorial moments of multiplicity distributions of
 inelastic and non-single-diffractive events obtained from the 2SGD fit at $\sqrt{s}$=8~TeV.} 
 {\begin{tabular}{@{}|c c c c c c c c c| @{}}  \toprule  
$\eta$ &  $C_2$ & $C_3$  & $C_4$ & $C_5$ &  $F_2$ & $F_3$  & $F_4$ & $F_5$ \\ 
 \hline 
\hline 
 \multicolumn{9}{|c|}{INEL}\\ 
 \hline-0.5 $< \eta <$ 0.5 &      1.784 $\pm$    0.019&       4.53 $\pm$     0.10&      14.47 $\pm$     0.46&      54.86 $\pm$     2.29&      1.638 $\pm$    0.018&       3.79 $\pm$     0.08 &     10.89 $\pm$     0.35 &     36.80 $\pm$     1.56 \\ \hline 
-1.0 $< \eta <$ 1.0 &      1.779 $\pm$    0.010&       4.42 $\pm$     0.05&      13.67 $\pm$     0.23&      49.62 $\pm$     1.09&      1.703 $\pm$    0.010&       4.03 $\pm$     0.05 &     11.76 $\pm$     0.20 &     40.07 $\pm$     0.89 \\ \hline 
-1.5 $< \eta <$ 1.5 &      1.800 $\pm$    0.008&       4.50 $\pm$     0.04&      13.93 $\pm$     0.17&      50.33 $\pm$     0.79&      1.747 $\pm$    0.007&       4.22 $\pm$     0.03 &     12.56 $\pm$     0.15 &     43.41 $\pm$     0.68 \\ \hline 
\hline 
 \multicolumn{9}{|c|}{INEL$>$0}\\ 
 \hline-1.0 $< \eta <$ 1.0 &      1.828 $\pm$    0.002&       4.68 $\pm$     0.01&      14.97 $\pm$     0.05&      56.50 $\pm$     0.26&      1.748 $\pm$    0.002&       4.26 $\pm$     0.01 &     12.88 $\pm$     0.04 &     45.64 $\pm$     0.21 \\ \hline 
\hline 
 \multicolumn{9}{|c|}{NSD}\\ 
 \hline-0.5 $< \eta <$ 0.5 &      1.766 $\pm$    0.015&       4.42 $\pm$     0.08&      13.93 $\pm$     0.37&      52.10 $\pm$     1.84&      1.622 $\pm$    0.014&       3.70 $\pm$     0.07 &     10.49 $\pm$     0.29 &     34.96 $\pm$     1.25 \\ \hline 
-1.0 $< \eta <$ 1.0 &      1.753 $\pm$    0.009&       4.28 $\pm$     0.04&      13.00 $\pm$     0.20&      46.35 $\pm$     0.94&      1.678 $\pm$    0.009&       3.90 $\pm$     0.04 &     11.19 $\pm$     0.18 &     37.42 $\pm$     0.77 \\ \hline 
-1.5 $< \eta <$ 1.5 &      1.751 $\pm$    0.009&       4.24 $\pm$     0.04&      12.69 $\pm$     0.18&      44.51 $\pm$     0.84&      1.700 $\pm$    0.009&       3.98 $\pm$     0.04 &     11.44 $\pm$     0.17 &     38.37 $\pm$     0.73 \\ \hline 
\hline  
 \multicolumn{9}{|c|}{INEL (Forward)}\\ 
 \hline-2.0 $< \eta <$ 2.0 &      1.802 $\pm$    0.003&       4.47 $\pm$     0.01&      13.45 $\pm$     0.06&      46.24 $\pm$     0.29&      1.762 $\pm$    0.003&       4.25 $\pm$     0.01 &     12.41 $\pm$     0.06 &     41.11 $\pm$     0.26 \\ \hline 
-2.4 $< \eta <$ 2.4 &      1.742 $\pm$    0.004&       4.15 $\pm$     0.02&      11.96 $\pm$     0.08&      39.24 $\pm$     0.34&      1.709 $\pm$    0.004&       3.98 $\pm$     0.02 &     11.17 $\pm$     0.07 &     35.52 $\pm$     0.31 \\ \hline 
-3.0 $< \eta <$ 3.0 &      1.699 $\pm$    0.004&       3.93 $\pm$     0.02&      11.02 $\pm$     0.07&      35.09 $\pm$     0.29&      1.673 $\pm$    0.004&       3.80 $\pm$     0.02 &     10.42 $\pm$     0.06 &     32.33 $\pm$     0.27 \\ \hline 
-3.4 $< \eta <$ 3.4 &      1.696 $\pm$    0.003&       3.91 $\pm$     0.01&      10.93 $\pm$     0.06&      34.67 $\pm$     0.26&      1.673 $\pm$    0.003&       3.80 $\pm$     0.01 &     10.39 $\pm$     0.06 &     32.20 $\pm$     0.24 \\ \hline 
-3.4 $< \eta <$ 5.0 &      1.666 $\pm$    0.003&       3.75 $\pm$     0.01&      10.18 $\pm$     0.06&      31.32 $\pm$     0.25&      1.647 $\pm$    0.003&       3.65 $\pm$     0.01 &      9.74 $\pm$     0.06 &     29.35 $\pm$     0.24 \\ \hline 
\hline 
 \multicolumn{9}{|c|}{INEL$>$0  (Forward)}\\ 
 \hline-2.0 $< \eta <$ 2.0 &      1.739 $\pm$    0.003&       4.13 $\pm$     0.01&      11.94 $\pm$     0.06&      39.41 $\pm$     0.25&      1.700 $\pm$    0.003&       3.94 $\pm$     0.01 &     11.02 $\pm$     0.05 &     35.05 $\pm$     0.22 \\ \hline 
-2.4 $< \eta <$ 2.4 &      1.694 $\pm$    0.004&       3.90 $\pm$     0.02&      10.89 $\pm$     0.07&      34.59 $\pm$     0.30&      1.663 $\pm$    0.004&       3.75 $\pm$     0.02 &     10.17 $\pm$     0.07 &     31.31 $\pm$     0.27 \\ \hline 
-3.0 $< \eta <$ 3.0 &      1.660 $\pm$    0.004&       3.73 $\pm$     0.02&      10.16 $\pm$     0.06&      31.45 $\pm$     0.26&      1.635 $\pm$    0.003&       3.61 $\pm$     0.02 &      9.61 $\pm$     0.06 &     28.97 $\pm$     0.25 \\ \hline 
-3.4 $< \eta <$ 3.4 &      1.657 $\pm$    0.003&       3.71 $\pm$     0.01&      10.05 $\pm$     0.06&      30.93 $\pm$     0.23&      1.634 $\pm$    0.003&       3.60 $\pm$     0.01 &      9.56 $\pm$     0.05 &     28.73 $\pm$     0.21 \\ \hline 
-3.4 $< \eta <$ 5.0 &      1.619 $\pm$    0.003&       3.52 $\pm$     0.01&       9.21 $\pm$     0.06&      27.34 $\pm$     0.23&      1.600 $\pm$    0.003&       3.43 $\pm$     0.01 &      8.81 $\pm$     0.05 &     25.62 $\pm$     0.21 \\ \hline 
\hline 
 \multicolumn{9}{|c|}{NSD  (Forward)}\\ 
 \hline2.4 $< \eta <$ 2.0 &      1.696 $\pm$    0.003&       3.89 $\pm$     0.01&      10.74 $\pm$     0.05&      33.73 $\pm$     0.21&      1.661 $\pm$    0.003&       3.71 $\pm$     0.01 &      9.93 $\pm$     0.05 &     30.07 $\pm$     0.18 \\ \hline 
-3.0 $< \eta <$ 2.4 &      1.674 $\pm$    0.003&       3.77 $\pm$     0.02&      10.22 $\pm$     0.06&      31.38 $\pm$     0.26&      1.644 $\pm$    0.003&       3.62 $\pm$     0.02 &      9.56 $\pm$     0.06 &     28.44 $\pm$     0.24 \\ \hline 
-3.4 $< \eta <$ 3.0 &      1.640 $\pm$    0.003&       3.61 $\pm$     0.01&       9.55 $\pm$     0.05&      28.65 $\pm$     0.22&      1.616 $\pm$    0.003&       3.49 $\pm$     0.01 &      9.04 $\pm$     0.05 &     26.43 $\pm$     0.20 \\ \hline 
-3.4 $< \eta <$ 3.4 &      1.634 $\pm$    0.003&       3.58 $\pm$     0.01&       9.44 $\pm$     0.05&      28.21 $\pm$     0.19&      1.612 $\pm$    0.003&       3.48 $\pm$     0.01 &      8.99 $\pm$     0.05 &     26.23 $\pm$     0.18 \\ \hline 
-3.4 $< \eta <$ 5.0 &      1.605 $\pm$    0.003&       3.43 $\pm$     0.01&       8.80 $\pm$     0.04&      25.52 $\pm$     0.16&      1.586 $\pm$    0.002&       3.34 $\pm$     0.01 &      8.42 $\pm$     0.04 &     23.94 $\pm$     0.15 \\ \hline 
\end{tabular}}
}
\end{center} 
 \end{table}  
  \begin{table}[pt]

 \begin{center}
 \scalebox{0.8}{ 
 \tbl{Normalized moments and normalized factorial moments of multiplicity distributions of
 inelastic and non-single-diffractive events obtained from the data at $\sqrt{s}$=7~TeV.~The uncertainties are combined statistical and systematic uncertainties.} 
 {\begin{tabular}{@{}|c c c c c c c c c| @{}}  \toprule 
$\eta$ &  $C_2$ & $C_3$  & $C_4$ & $C_5$ &  $F_2$ & $F_3$  & $F_4$ & $F_5$ \\ 
 \hline 
\hline 
 \multicolumn{9}{|c|}{INEL}\\ 
 \hline-0.5 $< \eta <$ 0.5 &      1.784 $\pm$    0.027&       4.54 $\pm$     0.14&      14.61 $\pm$     0.70&      55.93 $\pm$     3.57&      1.631 $\pm$    0.025&       3.77 $\pm$     0.12 &     10.90 $\pm$     0.53 &     37.08 $\pm$     2.41 \\ \hline 
-1.0 $< \eta <$ 1.0 &      1.804 $\pm$    0.015&       4.59 $\pm$     0.07&      14.58 $\pm$     0.32&      54.52 $\pm$     1.58&      1.724 $\pm$    0.014&       4.16 $\pm$     0.06 &     12.48 $\pm$     0.28 &     43.77 $\pm$     1.28 \\ \hline 
-1.5 $< \eta <$ 1.5 &      1.801 $\pm$    0.010&       4.55 $\pm$     0.05&      14.31 $\pm$     0.22&      52.90 $\pm$     1.11&      1.746 $\pm$    0.010&       4.26 $\pm$     0.04 &     12.87 $\pm$     0.20 &     45.51 $\pm$     0.97 \\ \hline 
\hline 
 \multicolumn{9}{|c|}{INEL$>$0}\\ 
 \hline-1.0 $< \eta <$ 1.0 &      1.804 $\pm$    0.011&       4.59 $\pm$     0.05&      14.57 $\pm$     0.25&      54.51 $\pm$     1.22&      1.724 $\pm$    0.011&       4.16 $\pm$     0.05 &     12.48 $\pm$     0.22 &     43.77 $\pm$     0.99 \\ \hline 
\hline 
 \multicolumn{9}{|c|}{NSD}\\ 
 \hline-0.5 $< \eta <$ 0.5 &      1.455 $\pm$    0.051&       2.88 $\pm$     0.20&       7.14 $\pm$     0.74&      21.06 $\pm$     2.88&      1.338 $\pm$    0.048&       2.39 $\pm$     0.17 &      5.33 $\pm$     0.56 &     13.96 $\pm$     1.94 \\ \hline 
-1.0 $< \eta <$ 1.0 &      1.770 $\pm$    0.025&       4.39 $\pm$     0.12&      13.62 $\pm$     0.55&      49.69 $\pm$     2.65&      1.691 $\pm$    0.024&       3.99 $\pm$     0.11 &     11.66 $\pm$     0.48 &     39.90 $\pm$     2.14 \\ \hline 
-1.5 $< \eta <$ 1.5 &      1.767 $\pm$    0.019&       4.36 $\pm$     0.09&      13.39 $\pm$     0.40&      48.26 $\pm$     1.93&      1.713 $\pm$    0.018&       4.08 $\pm$     0.08 &     12.04 $\pm$     0.37 &     41.51 $\pm$     1.67 \\ \hline 
\hline  
 \multicolumn{9}{|c|}{INEL (Forward)}\\ 
 \hline-2.0 $< \eta <$ 2.0 &      1.767 $\pm$    0.014&       4.28 $\pm$     0.08&      12.59 $\pm$     0.37&      42.11 $\pm$     1.73&      1.729 $\pm$    0.014&       4.08 $\pm$     0.07 &     11.62 $\pm$     0.34 &     37.44 $\pm$     1.55 \\ \hline 
-2.4 $< \eta <$ 2.4 &      1.752 $\pm$    0.019&       4.20 $\pm$     0.10&      12.19 $\pm$     0.47&      40.30 $\pm$     2.17&      1.720 $\pm$    0.019&       4.03 $\pm$     0.10 &     11.39 $\pm$     0.44 &     36.48 $\pm$     1.98 \\ \hline 
-3.0 $< \eta <$ 3.0 &      1.719 $\pm$    0.021&       4.03 $\pm$     0.11&      11.40 $\pm$     0.49&      36.68 $\pm$     2.21&      1.693 $\pm$    0.021&       3.89 $\pm$     0.11 &     10.78 $\pm$     0.47 &     33.79 $\pm$     2.05 \\ \hline 
-3.4 $< \eta <$ 3.4 &      1.690 $\pm$    0.021&       3.87 $\pm$     0.10&      10.70 $\pm$     0.47&      33.52 $\pm$     2.05&      1.666 $\pm$    0.021&       3.75 $\pm$     0.10 &     10.17 $\pm$     0.44 &     31.09 $\pm$     1.91 \\ \hline 
-3.4 $< \eta <$ 5.0 &      1.673 $\pm$    0.022&       3.77 $\pm$     0.11&      10.22 $\pm$     0.48&      31.30 $\pm$     2.08&      1.653 $\pm$    0.022&       3.67 $\pm$     0.11 &      9.77 $\pm$     0.46 &     29.29 $\pm$     1.96 \\ \hline 
\hline 
 \multicolumn{9}{|c|}{INEL$>$0  (Forward)}\\ 
 \hline-2.0 $< \eta <$ 2.0 &      1.704 $\pm$    0.011&       3.95 $\pm$     0.06&      11.11 $\pm$     0.26&      35.54 $\pm$     1.15&      1.667 $\pm$    0.011&       3.77 $\pm$     0.05 &     10.26 $\pm$     0.24 &     31.61 $\pm$     1.04 \\ \hline 
-2.4 $< \eta <$ 2.4 &      1.690 $\pm$    0.015&       3.88 $\pm$     0.07&      10.77 $\pm$     0.33&      34.03 $\pm$     1.46&      1.659 $\pm$    0.014&       3.72 $\pm$     0.07 &     10.07 $\pm$     0.31 &     30.81 $\pm$     1.33 \\ \hline 
-3.0 $< \eta <$ 3.0 &      1.674 $\pm$    0.016&       3.74 $\pm$     0.08&       9.97 $\pm$     0.34&      29.76 $\pm$     1.39&      1.648 $\pm$    0.016&       3.61 $\pm$     0.08 &      9.40 $\pm$     0.32 &     27.28 $\pm$     1.29 \\ \hline 
-3.4 $< \eta <$ 3.4 &      1.696 $\pm$    0.017&       3.88 $\pm$     0.08&      10.66 $\pm$     0.36&      33.26 $\pm$     1.60&      1.673 $\pm$    0.016&       3.76 $\pm$     0.08 &     10.13 $\pm$     0.35 &     30.85 $\pm$     1.49 \\ \hline 
-3.4 $< \eta <$ 5.0 &      1.700 $\pm$    0.018&       3.86 $\pm$     0.09&      10.48 $\pm$     0.40&      32.08 $\pm$     1.70&      1.680 $\pm$    0.018&       3.76 $\pm$     0.09 &     10.01 $\pm$     0.38 &     30.00 $\pm$     1.60 \\ \hline 
\hline 
 \multicolumn{9}{|c|}{NSD  (Forward)}\\ 
 \hline2.4 $< \eta <$ 2.0 &      1.712 $\pm$    0.012&       3.99 $\pm$     0.06&      11.24 $\pm$     0.28&      36.00 $\pm$     1.25&      1.675 $\pm$    0.012&       3.80 $\pm$     0.06 &     10.38 $\pm$     0.26 &     32.03 $\pm$     1.12 \\ \hline 
-3.0 $< \eta <$ 2.4 &      1.692 $\pm$    0.015&       3.88 $\pm$     0.08&      10.77 $\pm$     0.35&      33.96 $\pm$     1.55&      1.661 $\pm$    0.015&       3.73 $\pm$     0.08 &     10.07 $\pm$     0.33 &     30.76 $\pm$     1.41 \\ \hline 
-3.4 $< \eta <$ 3.0 &      1.662 $\pm$    0.017&       3.73 $\pm$     0.09&      10.10 $\pm$     0.37&      31.03 $\pm$     1.58&      1.637 $\pm$    0.017&       3.61 $\pm$     0.08 &      9.56 $\pm$     0.35 &     28.60 $\pm$     1.46 \\ \hline 
-3.4 $< \eta <$ 3.4 &      1.656 $\pm$    0.017&       3.69 $\pm$     0.08&       9.89 $\pm$     0.36&      30.02 $\pm$     1.53&      1.634 $\pm$    0.017&       3.58 $\pm$     0.08 &      9.40 $\pm$     0.34 &     27.86 $\pm$     1.43 \\ \hline 
-3.4 $< \eta <$ 5.0 &      1.616 $\pm$    0.018&       3.49 $\pm$     0.09&       9.02 $\pm$     0.36&      26.37 $\pm$     1.49&      1.597 $\pm$    0.018&       3.40 $\pm$     0.08 &      8.63 $\pm$     0.35 &     24.69 $\pm$     1.40 \\ \hline 
\end{tabular}}
}
\end{center} 
 \end{table}
   
\begin{table}[pt]
\begin{center}
 \scalebox{0.8}{ 
 \tbl{Normalized moments and normalized factorial moments of multiplicity distributions of
 inelastic and non-single-diffractive events obtained from 2SGD fit at $\sqrt{s}$=7~TeV.} 
 {\begin{tabular}{@{}|c c c c c c c c c| @{}}  \toprule 
 
$\eta$ &  $C_2$ & $C_3$  & $C_4$ & $C_5$ &  $F_2$ & $F_3$  & $F_4$ & $F_5$ \\ 
 \hline 
\hline 
 \multicolumn{9}{|c|}{INEL}\\ 
 \hline-0.5 $< \eta <$ 0.5 &      1.779 $\pm$    0.008&       4.50 $\pm$     0.04&      14.35 $\pm$     0.19&      54.36 $\pm$     0.95&      1.627 $\pm$    0.007&       3.74 $\pm$     0.03 &     10.67 $\pm$     0.14 &     35.87 $\pm$     0.64 \\ \hline 
-1.0 $< \eta <$ 1.0 &      1.800 $\pm$    0.003&       4.53 $\pm$     0.02&      14.24 $\pm$     0.07&      52.76 $\pm$     0.34&      1.719 $\pm$    0.003&       4.11 $\pm$     0.01 &     12.16 $\pm$     0.06 &     42.23 $\pm$     0.27 \\ \hline 
-1.5 $< \eta <$ 1.5 &      1.808 $\pm$    0.002&       4.56 $\pm$     0.01&      14.37 $\pm$     0.04&      53.49 $\pm$     0.21&      1.753 $\pm$    0.002&       4.27 $\pm$     0.01 &     12.91 $\pm$     0.04 &     45.99 $\pm$     0.18 \\ \hline 
\hline 
 \multicolumn{9}{|c|}{INEL$>$0}\\ 
 \hline-1.0 $< \eta <$ 1.0 &      1.801 $\pm$    0.003&       4.54 $\pm$     0.02&      14.25 $\pm$     0.07&      52.81 $\pm$     0.34&      1.720 $\pm$    0.003&       4.11 $\pm$     0.01 &     12.17 $\pm$     0.06 &     42.27 $\pm$     0.27 \\ \hline 
\hline 
 \multicolumn{9}{|c|}{NSD}\\ 
 \hline-0.5 $< \eta <$ 0.5 &      1.450 $\pm$    0.016&       2.86 $\pm$     0.06&       7.05 $\pm$     0.22&      20.63 $\pm$     0.83&      1.333 $\pm$    0.015&       2.37 $\pm$     0.05 &      5.25 $\pm$     0.16 &     13.63 $\pm$     0.55 \\ \hline 
-1.0 $< \eta <$ 1.0 &      1.755 $\pm$    0.006&       4.29 $\pm$     0.03&      13.06 $\pm$     0.14&      46.72 $\pm$     0.65&      1.677 $\pm$    0.006&       3.89 $\pm$     0.03 &     11.17 $\pm$     0.12 &     37.40 $\pm$     0.53 \\ \hline 
-1.5 $< \eta <$ 1.5 &      1.741 $\pm$    0.005&       4.20 $\pm$     0.02&      12.56 $\pm$     0.10&      44.27 $\pm$     0.45&      1.688 $\pm$    0.005&       3.93 $\pm$     0.02 &     11.29 $\pm$     0.09 &     38.04 $\pm$     0.39 \\ \hline 
\hline  
 \multicolumn{9}{|c|}{INEL (Forward)}\\ 
 \hline-2.0 $< \eta <$ 2.0 &      1.763 $\pm$    0.003&       4.27 $\pm$     0.01&      12.61 $\pm$     0.06&      42.58 $\pm$     0.26&      1.724 $\pm$    0.003&       4.07 $\pm$     0.01 &     11.64 $\pm$     0.05 &     37.89 $\pm$     0.23 \\ \hline 
-2.4 $< \eta <$ 2.4 &      1.741 $\pm$    0.003&       4.16 $\pm$     0.02&      12.10 $\pm$     0.07&      40.33 $\pm$     0.31&      1.709 $\pm$    0.003&       3.99 $\pm$     0.02 &     11.31 $\pm$     0.06 &     36.53 $\pm$     0.28 \\ \hline 
-3.0 $< \eta <$ 3.0 &      1.699 $\pm$    0.004&       3.94 $\pm$     0.02&      11.10 $\pm$     0.07&      35.83 $\pm$     0.30&      1.672 $\pm$    0.004&       3.80 $\pm$     0.02 &     10.49 $\pm$     0.07 &     33.02 $\pm$     0.27 \\ \hline 
-3.4 $< \eta <$ 3.4 &      1.675 $\pm$    0.004&       3.81 $\pm$     0.02&      10.55 $\pm$     0.06&      33.35 $\pm$     0.27&      1.651 $\pm$    0.004&       3.70 $\pm$     0.02 &     10.03 $\pm$     0.06 &     30.96 $\pm$     0.25 \\ \hline 
-3.4 $< \eta <$ 5.0 &      1.647 $\pm$    0.004&       3.66 $\pm$     0.02&       9.88 $\pm$     0.06&      30.41 $\pm$     0.25&      1.626 $\pm$    0.004&       3.56 $\pm$     0.01 &      9.44 $\pm$     0.06 &     28.48 $\pm$     0.24 \\ \hline 
\hline 
 \multicolumn{9}{|c|}{INEL$>$0  (Forward)}\\ 
 \hline-2.0 $< \eta <$ 2.0 &      1.701 $\pm$    0.002&       3.95 $\pm$     0.01&      11.16 $\pm$     0.05&      36.07 $\pm$     0.22&      1.664 $\pm$    0.002&       3.76 $\pm$     0.01 &     10.30 $\pm$     0.05 &     32.10 $\pm$     0.20 \\ \hline 
-2.4 $< \eta <$ 2.4 &      1.682 $\pm$    0.003&       3.85 $\pm$     0.01&      10.75 $\pm$     0.06&      34.34 $\pm$     0.26&      1.651 $\pm$    0.003&       3.70 $\pm$     0.01 &     10.05 $\pm$     0.06 &     31.10 $\pm$     0.24 \\ \hline 
-3.0 $< \eta <$ 3.0 &      1.664 $\pm$    0.003&       3.71 $\pm$     0.01&       9.94 $\pm$     0.05&      29.99 $\pm$     0.22&      1.638 $\pm$    0.003&       3.59 $\pm$     0.01 &      9.38 $\pm$     0.05 &     27.51 $\pm$     0.20 \\ \hline 
-3.4 $< \eta <$ 3.4 &      1.649 $\pm$    0.004&       3.67 $\pm$     0.02&       9.92 $\pm$     0.06&      30.60 $\pm$     0.27&      1.626 $\pm$    0.004&       3.56 $\pm$     0.02 &      9.43 $\pm$     0.06 &     28.41 $\pm$     0.25 \\ \hline 
-3.4 $< \eta <$ 5.0 &      1.637 $\pm$    0.003&       3.61 $\pm$     0.01&       9.65 $\pm$     0.05&      29.38 $\pm$     0.21&      1.617 $\pm$    0.003&       3.51 $\pm$     0.01 &      9.23 $\pm$     0.05 &     27.52 $\pm$     0.20 \\ \hline 
\hline 
 \multicolumn{9}{|c|}{NSD  (Forward)}\\ 
 \hline2.4 $< \eta <$ 2.0 &      1.709 $\pm$    0.002&       3.98 $\pm$     0.01&      11.29 $\pm$     0.05&      36.58 $\pm$     0.22&      1.672 $\pm$    0.002&       3.80 $\pm$     0.01 &     10.43 $\pm$     0.05 &     32.57 $\pm$     0.20 \\ \hline 
-3.0 $< \eta <$ 2.4 &      1.687 $\pm$    0.003&       3.87 $\pm$     0.01&      10.81 $\pm$     0.06&      34.54 $\pm$     0.26&      1.656 $\pm$    0.003&       3.72 $\pm$     0.01 &     10.11 $\pm$     0.06 &     31.30 $\pm$     0.24 \\ \hline 
-3.4 $< \eta <$ 3.0 &      1.652 $\pm$    0.003&       3.70 $\pm$     0.02&      10.05 $\pm$     0.06&      31.25 $\pm$     0.26&      1.627 $\pm$    0.003&       3.58 $\pm$     0.01 &      9.51 $\pm$     0.06 &     28.82 $\pm$     0.24 \\ \hline 
-3.4 $< \eta <$ 3.4 &      1.644 $\pm$    0.004&       3.65 $\pm$     0.02&       9.84 $\pm$     0.06&      30.24 $\pm$     0.26&      1.622 $\pm$    0.004&       3.54 $\pm$     0.01 &      9.36 $\pm$     0.06 &     28.09 $\pm$     0.24 \\ \hline 
0.0 $< \eta <$ 5.0 &      1.603 $\pm$    0.004&       3.45 $\pm$     0.02&       9.02 $\pm$     0.06&      26.87 $\pm$     0.25&      1.584 $\pm$    0.004&       3.36 $\pm$     0.02 &      8.62 $\pm$     0.06 &     25.19 $\pm$     0.24 \\ \hline 
\end{tabular}} 
}
\end{center}
 \end{table}

 \begin{table}[pt] 
 \begin{center}
 \scalebox{0.8}{
 \tbl{Normalized moments and normalized factorial moments of multiplicity distributions of
 inelastic and non-single-diffractive events obtained from the data at $\sqrt{s}$=2.76~TeV.~The uncertainties are combined statistical and systematic uncertainties} 
 {\begin{tabular}{@{}|c c c c c c c c c| @{}}  \toprule 
 $\eta$ &  $C_2$ & $C_3$  & $C_4$ & $C_5$ &  $F_2$ & $F_3$  & $F_4$ & $F_5$ \\ 
 \hline 
\hline 
 \multicolumn{9}{|c|}{INEL}\\ 
 \hline-0.5 $< \eta <$ 0.5 &      1.710 $\pm$    0.023&       4.15 $\pm$     0.12&      12.77 $\pm$     0.54&      46.96 $\pm$     2.71&      1.525 $\pm$    0.021&       3.26 $\pm$     0.09 &      8.76 $\pm$     0.39 &     27.72 $\pm$     1.66 \\ \hline 
-1.0 $< \eta <$ 1.0 &      1.723 $\pm$    0.016&       4.16 $\pm$     0.08&      12.55 $\pm$     0.35&      44.74 $\pm$     1.65&      1.624 $\pm$    0.015&       3.66 $\pm$     0.07 &     10.25 $\pm$     0.29 &     33.63 $\pm$     1.27 \\ \hline 
-1.5 $< \eta <$ 1.5 &      1.713 $\pm$    0.015&       4.08 $\pm$     0.07&      12.06 $\pm$     0.33&      41.81 $\pm$     1.58&      1.645 $\pm$    0.014&       3.73 $\pm$     0.07 &     10.47 $\pm$     0.29 &     34.21 $\pm$     1.32 \\ \hline 
\hline 
  \multicolumn{9}{|c|}{INEL$>0$}\\ 
 \hline-1.0 $< \eta <$ 1.0 &      1.723 $\pm$    0.012&       4.15 $\pm$     0.06&      12.54 $\pm$     0.26&      44.73 $\pm$     1.26&      1.624 $\pm$    0.012&       3.66 $\pm$     0.05 &     10.25 $\pm$     0.22 &     33.62 $\pm$     0.96 \\ \hline 
\hline 
  \multicolumn{9}{|c|}{NSD}\\ 
 \hline-0.5 $< \eta <$ 0.5 &      1.687 $\pm$    0.051&       4.01 $\pm$     0.25&      12.09 $\pm$     1.11&      43.48 $\pm$     5.31&      1.506 $\pm$    0.047&       3.16 $\pm$     0.20 &      8.29 $\pm$     0.78 &     25.68 $\pm$     3.21 \\ \hline 
-1.0 $< \eta <$ 1.0 &      1.688 $\pm$    0.036&       3.96 $\pm$     0.17&      11.60 $\pm$     0.73&      40.11 $\pm$     3.31&      1.591 $\pm$    0.034&       3.49 $\pm$     0.15 &      9.48 $\pm$     0.60 &     30.16 $\pm$     2.52 \\ \hline 
-1.5 $< \eta <$ 1.5 &      1.675 $\pm$    0.031&       3.87 $\pm$     0.14&      11.08 $\pm$     0.60&      37.18 $\pm$     2.70&      1.609 $\pm$    0.030&       3.54 $\pm$     0.13 &      9.62 $\pm$     0.53 &     30.43 $\pm$     2.23 \\ \hline 
\end{tabular}}
}
\end{center} 
 \end{table}

\begin{table}[pt]

\begin{center}
 \scalebox{0.8}{ 
 \tbl{Normalized moments and normalized factorial moments of multiplicity distributions of
 inelastic and non-single-diffractive events obtained from the 2SGD fit at $\sqrt{s}$=2.76~TeV.} 
 {\begin{tabular}{@{}|c c c c c c c c c| @{}}  \toprule 
 
$\eta$ &  $C_2$ & $C_3$  & $C_4$ & $C_5$ &  $F_2$ & $F_3$  & $F_4$ & $F_5$ \\ 
 \hline 
\hline 
 \multicolumn{9}{|c|}{INEL}\\ 
 \hline-0.5 $< \eta <$ 0.5 &      1.711 $\pm$    0.008&       4.15 $\pm$     0.04&      12.73 $\pm$     0.18&      46.73 $\pm$     0.87&      1.525 $\pm$    0.008&       3.26 $\pm$     0.03 &      8.71 $\pm$     0.13 &     27.52 $\pm$     0.52 \\ \hline 
-1.0 $< \eta <$ 1.0 &      1.732 $\pm$    0.005&       4.18 $\pm$     0.02&      12.63 $\pm$     0.09&      45.14 $\pm$     0.44&      1.631 $\pm$    0.004&       3.68 $\pm$     0.02 &     10.30 $\pm$     0.08 &     33.87 $\pm$     0.34 \\ \hline 
-1.5 $< \eta <$ 1.5 &      1.703 $\pm$    0.004&       4.02 $\pm$     0.02&      11.84 $\pm$     0.08&      41.18 $\pm$     0.35&      1.634 $\pm$    0.004&       3.68 $\pm$     0.02 &     10.28 $\pm$     0.07 &     33.72 $\pm$     0.29 \\ \hline 
\hline 
  \multicolumn{9}{|c|}{INEL$>0$}\\ 
 \hline-1.0 $< \eta <$ 1.0 &      1.729 $\pm$    0.005&       4.17 $\pm$     0.02&      12.57 $\pm$     0.09&      44.91 $\pm$     0.44&      1.629 $\pm$    0.004&       3.67 $\pm$     0.02 &     10.25 $\pm$     0.08 &     33.70 $\pm$     0.33 \\ \hline 
\hline 
  \multicolumn{9}{|c|}{NSD}\\ 
 \hline-0.5 $< \eta <$ 0.5 &      1.683 $\pm$    0.019&       3.99 $\pm$     0.09&      11.95 $\pm$     0.41&      42.73 $\pm$     1.95&      1.502 $\pm$    0.018&       3.14 $\pm$     0.07 &      8.19 $\pm$     0.29 &     25.18 $\pm$     1.17 \\ \hline 
-1.0 $< \eta <$ 1.0 &      1.681 $\pm$    0.014&       3.91 $\pm$     0.06&      11.35 $\pm$     0.26&      38.90 $\pm$     1.17&      1.585 $\pm$    0.013&       3.45 $\pm$     0.06 &      9.26 $\pm$     0.22 &     29.19 $\pm$     0.89 \\ \hline 
-1.5 $< \eta <$ 1.5 &      1.657 $\pm$    0.010&       3.76 $\pm$     0.04&      10.61 $\pm$     0.18&      35.17 $\pm$     0.77&      1.591 $\pm$    0.009&       3.45 $\pm$     0.04 &      9.22 $\pm$     0.16 &     28.78 $\pm$     0.64 \\ \hline 
\end{tabular}}
}
\end{center} 
 \end{table}

\begin{figure}[th]
\centerline{\includegraphics[scale=0.70]{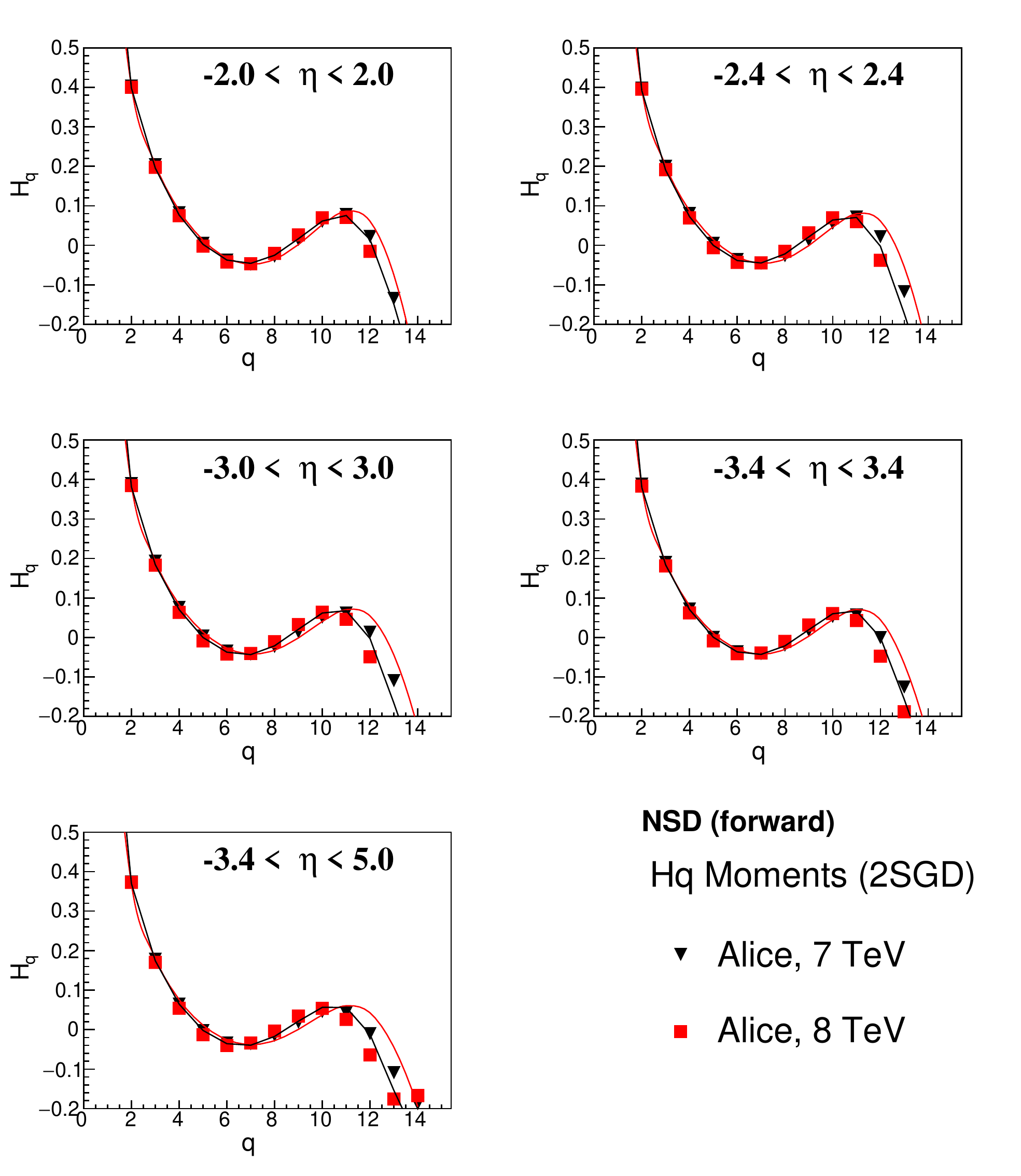}}
\caption{$H_{q}$ moments of charged multiplicity of NSD events 8 and 7~TeV in forward region of the ALICE detector at the LHC.~The distributions are derived from 2SGD fits to the data as represented in the legends.}
\label{figHq2SG}
\end{figure}

\begin{figure}[th]
\centerline{\includegraphics[scale=0.70]{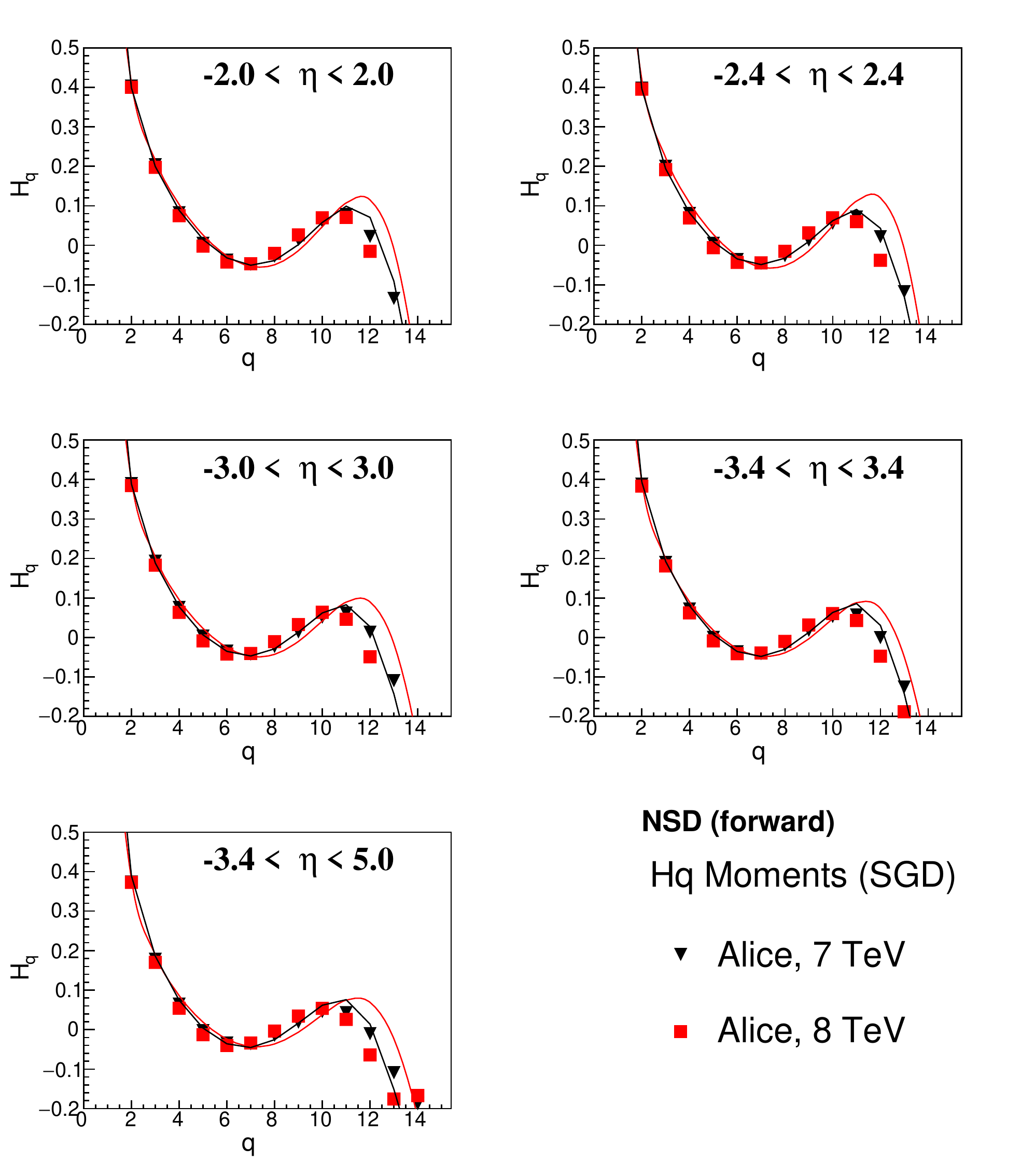}}
\caption{$H_{q}$ moments of charged multiplicity of NSD events 8 and 7~TeV in forward region of the ALICE detector at the LHC.~The distributions are derived from SGD fits to the data as represented in the legends.}
\label{figHqSG}
\end{figure} 

Higher-order moments and the cumulants are the precise tools for exploring the possibility of finding correlations amongst the charged particles produced in collisions [\refcite{a20}].~The deviation w.r.t. independent and uncorrelated production of particles can be measured well using the factorial moments, $F_q$ [\refcite{a21}].
Figure \ref{figFq} shows the normalized factorial moments $F_{q}$ for $q$=2,\,3,\,4,\,5 for $\sqrt{s}$=\,8,\,7 and 2.76~TeV for inelastic and NSD collisions in three pseudorapidity intervals, $|\eta|<$ 0.5, $|\eta|<$ 1.0 and $|\eta|<$ 1.5.~The values obtained by fitting the 2SGD and SGD are shown in comprison to the experimental values.~The values calculated from 2SGD reproduce the experimental values nearly accurately.~It may be observed that values of moments $F_{2}$,\,$F_{3}$ remain roughly constant with energy while higher moments $F_{4}$,\,$F_{5}$ show a linear increase with increasing collision energy.~Moreover, the increase becomes more dominant for larger rapidity windows.~This clearly dictates the violation of KNO scaling at high energies.~Values of $C_{q}$ and $F_{q}$ moments at all energies and in all pseudorapidity windows are given in tables~5-10.~The results are obtained only for $\sqrt{s}$= 8 and 7~TeV in the forward region, as no data are available for 2.76~TeV.

Shape of the multiplicity distribution is also analysed in terms of the $H{_q}$ moments.~It was established that a special oscillation pattern for the ratio of cumulants to factorial moments $H{_q}$ = $K_{q}/F_{q}$ occurs.~It was predicted that the oscillations take place with first minimum occuring around $q_{min}\sim$5.~The predictions were derived from the inverse QCD anomalous dimension and the QCD running coupling constant determined from the number of colours as $N_c$=3.~Details can be found in references [\refcite{a22},~\refcite{a23},~\refcite{a24},~\refcite{a25}].~When we analysed hadronic multiplicities in $e^+e^−$ and $\overline{p}p$/$pp$ experiments [\refcite{a26}] using SGD, similar oscillations in the $H{_q}$ moments were observed.~In the present work, we study the behaviour of $H{_q}$ moments to check whether the multiplicity distributions in different pseudorapidity windows and also in the forward region follow the same trends.~Also, the shape of the multiplicity distribution analyzed in terms of the $H_{q}$ moments was found to reveal quasi-oscillations.~In the next-to-next-to-leading-log-approximation~(NNLLA) of perturbative QCD, a negative first minimum is expected near $q\sim$5 and quasi-oscillations about zero are expected for larger values of $q$.~We analyse both inelastic and NSD events using single as well as two${\text -}$component shifted Gompertz distribution. 

\begin{figure}[th]
\centerline{\includegraphics[scale=0.40]{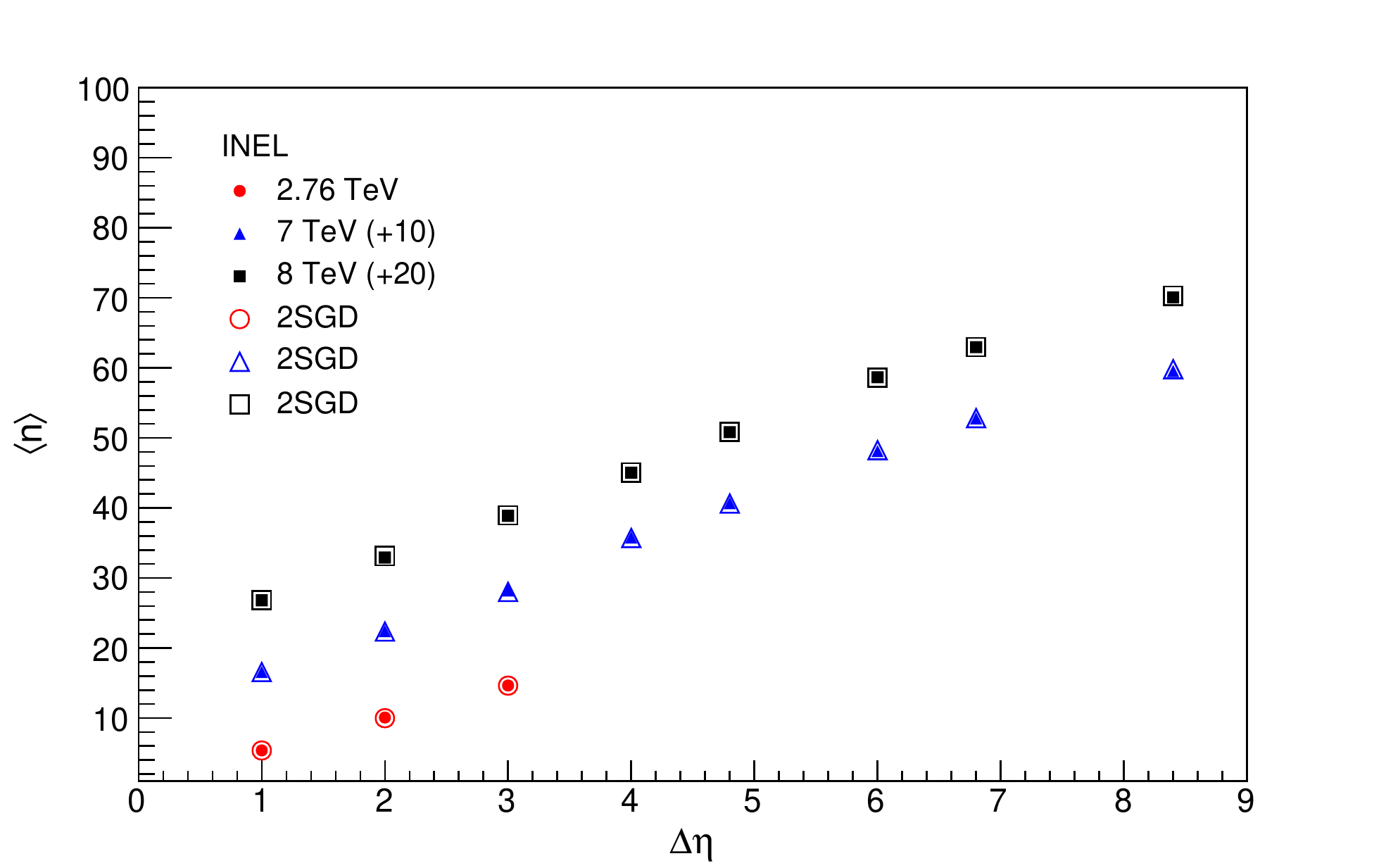}\includegraphics[scale=0.40]{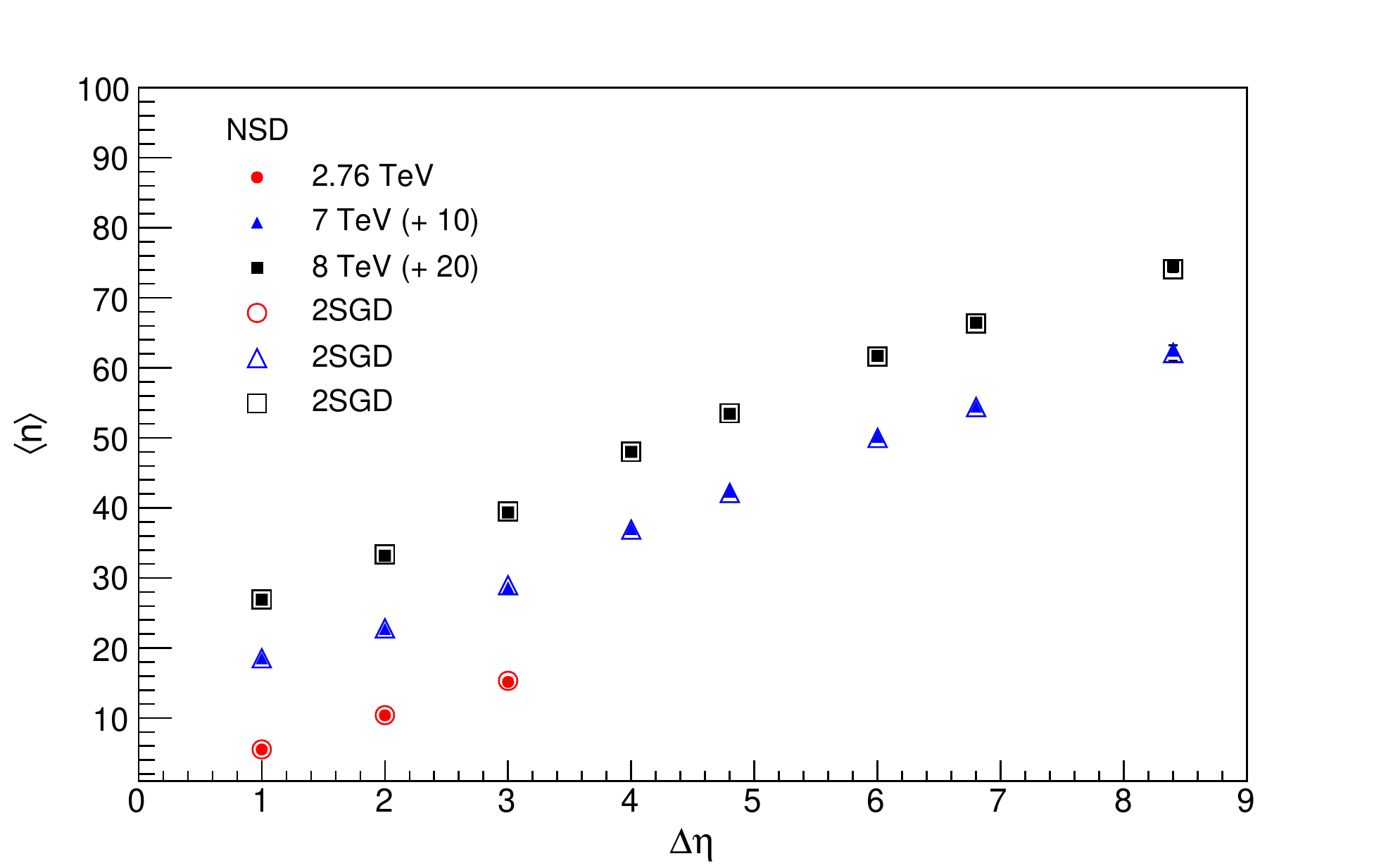}}
\caption{Dependence of average charged multiplicity on $\Delta\eta$ at different energies in inelastic and NSD events from the ALICE detector at the LHC.}
\label{figAvgn}
\end{figure}

We present the first results on the $H_{q}$ moments.~Figures \ref{figHq2SG} and \ref{figHqSG} show the $H{_q}$ moments versus $q$ values calculated from the data and compared with 2SGD and SGD distributions in different $\eta$ bins, for the inelastic and NSD events.~We observe that 2SGD best describes the data with minimum around $q\sim$􏰁7.~We find that the dependence of $H{_q}$ on $q$ is very similar in all the $\eta$ windows with a steep descent to a minimum value around $q$=6-7.~In the forward region for wider $\eta$-windows ($|\eta|<$2, $|\eta|<$2.4, $|\eta|<$3, $|\eta|<$3.4 and -3.4$<\eta<$5) the amplitude of the first minimum and the first maximum is roughly constant.~Quasi${\text -}$oscillations about zero are expected for larger values of $q$.~These observations confirm the predictions from Quantum Chromodynamics and also the next-to-next-to-leading logarithm approximation (NNLLA) of perturbative QCD. 
 \begin{table}[pt] 
 
 \begin{center}
 \scalebox{0.8}{ 
 \tbl{Comparison of average charged multiplicity obtained from the data with 2SGD fit values.} 
    {\begin{tabular}{@{}|c|c  c c c  c  c|@{}}  \toprule 
 
 $\eta$ & \multicolumn{2}{c}{$\langle{n}\rangle \,\,$8~TeV} & \multicolumn{2}{c}{$\langle{n}\rangle \,\,$7~TeV} & \multicolumn{2}{c|}{$\langle{n}\rangle \,\,$2.76~TeV} \\ 
  &       &      &      &      &      &      \\       
  &  Exp. & 2SGD & Exp. & 2SGD & Exp. & 2SGD  \\

 \hline\hline 
 \multicolumn{7}{|c|}{INEL}\\ 
 \hline-0.5 $< \eta <$ 0.5 &       6.81 $\pm$     0.15 &       6.83 $\pm$     0.06 &           6.57 $\pm$     0.06 &       6.57 $\pm$     0.02 &       5.38 $\pm$     0.05 &       5.37 $\pm$     0.02  \\ 
 \hline 
-1.0 $< \eta <$ 1.0 &      12.91 $\pm$     0.21 &      13.14 $\pm$     0.06 &          12.42 $\pm$     0.06 &      12.35 $\pm$     0.02 &      10.07 $\pm$     0.06 &       9.98 $\pm$     0.02  \\ 
 \hline 
-1.5 $< \eta <$ 1.5 &      18.88 $\pm$     0.25 &      18.95 $\pm$     0.06 &          18.21 $\pm$     0.07 &      18.01 $\pm$     0.01 &      14.65 $\pm$     0.10 &      14.63 $\pm$     0.03  \\ 
 \hline 
\hline 
 \multicolumn{7}{|c|}{INEL $ >$ 0}\\ 
 \hline-1.0 $< \eta <$ 1.0 &      12.91 $\pm$     0.05 &      12.65 $\pm$     0.01 &          12.42 $\pm$     0.06 &      12.35 $\pm$     0.02 &      10.07 $\pm$     0.06 &       9.99 $\pm$     0.02  \\ 
 \hline 
\hline 
 \multicolumn{7}{|c|}{NSD}\\ 
 \hline-0.5 $< \eta <$ 0.5 &       6.90 $\pm$     0.14 &       6.93 $\pm$     0.05 &           8.53 $\pm$     0.16 &       8.54 $\pm$     0.07 &       5.51 $\pm$     0.12 &       5.52 $\pm$     0.05  \\ 
 \hline 
-1.0 $< \eta <$ 1.0 &      13.18 $\pm$     0.19 &      13.37 $\pm$     0.06 &          12.74 $\pm$     0.13 &      12.81 $\pm$     0.04 &      10.39 $\pm$     0.17 &      10.41 $\pm$     0.07  \\ 
 \hline 
-1.5 $< \eta <$ 1.5 &      19.35 $\pm$     0.23 &      19.49 $\pm$     0.08 &          18.67 $\pm$     0.15 &      18.97 $\pm$     0.04 &      15.15 $\pm$     0.22 &      15.32 $\pm$     0.07  \\ 
 \hline 
\hline \hline 
 \multicolumn{7}{|c|}{INEL (Forward)}\\ 
 \hline-2.0 $< \eta <$ 2.0 &          25.03 $\pm$     0.18 &      25.06 $\pm$     0.03 &      25.78 $\pm$     0.20 &      25.71 $\pm$     0.03  &   &  \\ 
 \hline 
-2.4 $< \eta <$ 2.4 &          30.82 $\pm$     0.35 &      30.84 $\pm$     0.06 &      30.68 $\pm$     0.32 &      30.63 $\pm$     0.05  &   &  \\ 
 \hline 
-3.0 $< \eta <$ 3.0 &          38.66 $\pm$     0.47 &      38.62 $\pm$     0.07 &      38.12 $\pm$     0.44 &      38.25 $\pm$     0.07  &   &  \\ 
 \hline 
-3.4 $< \eta <$ 3.4 &          42.96 $\pm$     0.51 &      42.97 $\pm$     0.07 &      42.78 $\pm$     0.51 &      42.82 $\pm$     0.08  &   &  \\ 
 \hline 
-3.4 $< \eta <$ 5.0 &          50.08 $\pm$     0.68 &      50.29 $\pm$     0.09 &      49.56 $\pm$     0.66 &      49.81 $\pm$     0.09  &   &  \\ 
 \hline 
\hline 
 \multicolumn{7}{|c|}{INEL $ >$ 0  (Forward)}\\ 
 \hline-2.0 $< \eta <$ 2.0 &          26.12 $\pm$     0.19 &      26.12 $\pm$     0.04 &      26.97 $\pm$     0.21 &      26.87 $\pm$     0.04  &   &  \\ 
 \hline 
-2.4 $< \eta <$ 2.4 &          31.88 $\pm$     0.36 &      31.88 $\pm$     0.06 &      32.10 $\pm$     0.33 &      31.98 $\pm$     0.05  &   &  \\ 
 \hline 
-3.0 $< \eta <$ 3.0 &          39.87 $\pm$     0.49 &      39.78 $\pm$     0.07 &      38.84 $\pm$     0.45 &      38.85 $\pm$     0.06  &   &  \\ 
 \hline 
-3.4 $< \eta <$ 3.4 &          44.36 $\pm$     0.53 &      44.32 $\pm$     0.07 &      42.99 $\pm$     0.51 &      43.90 $\pm$     0.08  &   &  \\ 
 \hline 
-3.4 $< \eta <$ 5.0 &          52.47 $\pm$     0.72 &      52.19 $\pm$     0.10 &      49.11 $\pm$     0.66 &      50.38 $\pm$     0.08  &   &  \\ 
 \hline 
\hline 
 \multicolumn{7}{|c|}{NSD  (Forward)}\\ 
 \hline-2.0 $< \eta <$ 2.0 &          28.00 $\pm$     0.21 &      28.01 $\pm$     0.04 &      27.01 $\pm$     0.21 &      26.91 $\pm$     0.04  &   &  \\ 
 \hline 
-2.4 $< \eta <$ 2.4 &          33.44 $\pm$     0.39 &      33.51 $\pm$     0.06 &      32.31 $\pm$     0.34 &      32.12 $\pm$     0.05  &   &  \\ 
 \hline 
-3.0 $< \eta <$ 3.0 &          41.70 $\pm$     0.52 &      41.61 $\pm$     0.07 &      40.16 $\pm$     0.47 &      40.00 $\pm$     0.07  &   &  \\ 
 \hline 
-3.4 $< \eta <$ 3.4 &          46.45 $\pm$     0.57 &      46.35 $\pm$     0.07 &      44.47 $\pm$     0.54 &      44.40 $\pm$     0.08  &   &  \\ 
 \hline 
-3.4 $< \eta <$ 5.0 &          54.47 $\pm$     0.76 &      54.09 $\pm$     0.08 &      52.46 $\pm$     0.71 &      52.12 $\pm$     0.11  &   &  \\ 
 \hline
 
\end{tabular}}
}
\end{center} 
 \end{table} 
Finally the multiplicity distributions are also analysed in terms of the average multiplicities.~Table~11 shows the values of average charged particle multiplicity values calculated from the data and from the 2SGD fitted distributions at different energies in different pseudorapidity bins.~Figure~\ref{figAvgn} shows the average charged multiplicity dependence on the size of rapidity window, $\Delta\eta$~(-0.5$<\eta<0.5$ denotes $\Delta\eta=1$ and so on) at three energies both for the data and the values from 2SGD fits.~It may be observed that at each energy, the $\langle{n}\rangle$ increases almost linearly with $\Delta\eta$.~However, a comparison shows that the variation of $\langle{n}\rangle$ as a function of $|\Delta\eta|$ is nearly the same for c.m.s energy from 2.76 to 8~TeV.

\section{Conclusions}
Analysis of multiplicity distributions at $\sqrt{s}$=8,\,7 and 2.76~TeV, in  pseudorapidity windows varying in size, obtained by the ALICE experiment at the LHC is performed to compare the data with the shifted Gompertz distribution Eq.(\ref{SGD}) and its modified form Eq.(\ref{2SGD}).~It is pertinent to mention that the ALICE collaboration, in one of its papers [\refcite{ALICE1}] has analysed the data by using NBD.~The presented analysis is to study the applicability of the shifted Gompertz distribution~(SGD).~The analysis shows that the SGD explains the data very well in larger $\eta$ windows.~However, in smaller $\eta$ windows (typically $|\eta|$ = 0.5,\,1.0,\,1.5), the SGD fits deviate from the data.~While studying the multiplicity distributions of $\overline{p}p$ collisions at c.m.s energies of 200${\text -}$900~GeV, such a behaviour was also observed with negative binomial fits~[10, 12].~A possible explanation of this effect was suggested by C. Fuglesang [13] proposing the presence of a substructure in the interactions.~To correct this violation, it was proposed that the multiplicity distribution consists of contributions from two different types of events; soft and semi${\text -}$hard events.~Each component could be separately described by a negative binomial function.~A superposition of the two component distribution describes the total multiplicity distribution.~Using this approach, we modified the $P(n)$ versus $n$ dependence in terms of two-component shifted Gompertz distribution~(2SGD) as in Eq.(\ref{2SGD}).~Fitting the data with 2SGD resulted in manifold reduction in the $\chi^{2}/ndf$ values and the distributions at all energies described the data very well, with $p{\text -}$values corresponding to $CL>$ 0.1\%.

Shape of the charged${\text -}$particle multiplicity distribution depends upon c.m.s energy available for particle production and the type of colliding particles.~The normalized factorial moments define the correlations amongst particles being produced.~The multiplicity distribution obeys conventional Poisson distribution if there is no correlation between the particles produced, in which case the factorial moments are of the order of unity.~The results confirm the change of slope in probability distribution in nearly all pseudorapidity intervals observed at 8,\,7 and 2.76~TeV.~Broadening of the distribution beyond the expectation of Poisson distribution, indicates the presence of particle correlations, with the $F_{q}$ values becoming greater than unity.~If the particles are anti-correlated, the distribution becomes narrower reducing the $F_{q}$ values to less than unity. 

We find that the moments $C_{q}$ and $F_{q}$ are greater than unity for every energy and every pseudorapidity window.~The moments tend to increase with c.m.s energy.~The increase is more significant for higher moments with $q>$3.~This emphasises the presence of strong positive correlations in particle production.~The values amongst different $\eta$ bins, however are roughly constant within the error limits.

Study of $H_{q}$ moments shows that dependence of $H_{q}$ on $q$ is very similar in all the $\eta$ bins with a negative minimum value around $q\sim$6$-$7 for both inelastic and non-single-diffractive events.~We observe a steep descent of $H_q$ to the first minimum for all $\eta$ wondows.~In the forward region for wider $\eta$-windows ($|\eta|\geq$2) the amplitude of first  minimum and first maximum is roughly constant.~Quasi-oscillations about zero are expected for larger values of $q$.~These observations confirm the predictions from Quantum Chromodynamics and also the next-to-next-to-leading logarithm approximation (NNLLA) of perturbative QCD [\refcite{a22},~\refcite{a23},~\refcite{a24},~\refcite{a25}].~For 2.76~TeV collisions, as the data are not available in the forward region $|\eta|>$2.0, the moments are not reported and it is not possible to study this dependence for the pseudorapidity range 2.0$<|\eta|<$5.0 in this case.
 
The average charged multiplicity $\langle{n}\rangle$ for both the INEL and NSD events increases with the pseudorapidity window.~The $\langle{n}\rangle$ values from the data agree very well with the 2SGD values.~Applicability of two-component shifted Gompertz distribution is in very good agreement with the data.~The predictions made by shifted Gompertz distribution, serve as a good test of the validity of this proposed distribution for the ALICE data at three energies.
~In conclusion, a double-SGD function provides a precise description of the entire set of multiplicity distributions measured in this experiment.

\section*{Acknowledgements}
The author R.~Aggarwal is grateful to the Department of Science and Technology, Government of India, for the INSPIRE faculty grant.

\end{document}